\documentclass[journal]{IEEEtran}

\usepackage{soul}
\usepackage{color}
\usepackage{graphicx}
\usepackage{amsmath}
\usepackage{cite}
\usepackage{subfig}
\usepackage{multirow}
\usepackage{booktabs}
\usepackage{hyperref}

\begin{document}
%
\title{Transfer Learning from Speech Synthesis to Voice Conversion with Non-Parallel Training Data}
%
%
%

\author{Mingyang~Zhang,~\IEEEmembership{Member,~IEEE,}
		Yi~Zhou,~\IEEEmembership{Student~Member,~IEEE,}
		Li~Zhao,
		and~Haizhou~Li,~\IEEEmembership{Fellow,~IEEE}
\thanks{Mingyang Zhang and Li Zhao are with the School of Information Science and Engineering, Southeast University, Nanjing, China (email: zhangmy@seu.edu.cn; zhaoli@seu.edu.cn)}
\thanks{Yi Zhou and Haizhou Li are with the Department of Electrical and Computer Engineering, National University of Singapore, Singapore (email: yi.zhou@u.nus.edu; haizhou.li@nus.edu.sg)}
}

%
%

\markboth{IEEE/ACM Transactions on Audio, Speech, and Language Processing}%
{Zhang \MakeLowercase{\textit{et al.}}: Transfer Learning from Speech Synthesis to Voice Conversion with Non-Parallel Training Data}
%



\maketitle

\begin{abstract}
This paper presents a novel framework to build a voice conversion (VC) system by learning from a text-to-speech (TTS) synthesis system, that is called TTS-VC transfer learning. We first develop a multi-speaker speech synthesis system with sequence-to-sequence encoder-decoder architecture, where the encoder extracts robust linguistic representations of text, and the decoder, conditioned on target speaker embedding, takes the context vectors and the attention recurrent network cell output to generate target acoustic features. We take advantage of the fact that TTS system maps input text to speaker independent context vectors, and reuse such a mapping to supervise the training of latent representations of an encoder-decoder voice conversion system. In the voice conversion system, the encoder takes speech instead of text as input, while the decoder is functionally similar to TTS decoder. 
As we condition the decoder on speaker embedding, the system can be trained on non-parallel data for any-to-any voice conversion. During voice conversion training, we present both text and speech to speech synthesis and voice conversion networks respectively. At run-time, the voice conversion network uses its own encoder-decoder architecture. Experiments show that the proposed approach outperforms two competitive voice conversion baselines consistently, namely phonetic posteriorgram and variational autoencoder methods, in terms of speech quality, naturalness, and speaker similarity.

\end{abstract}


%
\IEEEpeerreviewmaketitle

\section{Introduction}
\label{sec:intro}
%
%
%
%
\IEEEPARstart{V}{oice} Conversion (VC) takes speech of the source speaker as input and generates speech that sounds from a target speaker while maintaining the linguistic content. It is an enabling technology for many innovative applications, such as personalized or expressive speech synthesis~\cite{vc_tts, 1643640}, speech enhancement~\cite{TodaEnhence}, normalization of impaired speech~\cite{NAKAMURA2012134, Chen2019}, speech-to-singing conversion~\cite{new2010voice, 4393001} and dubbing of movies and games. In general, voice conversion techniques are broadly grouped into parallel and non-parallel methods according to their use of training data~\cite{mouchtaris2007conditional}. Parallel voice conversion requires source-target pair of speech samples of the same linguistic content, while non-parallel voice conversion is trained on unpaired speech data.

Parallel voice conversion can be formulated as a regression problem where a mapping function is estimated between source and target spectral features. Successful techniques include Gaussian mixture model (GMM)~\cite{Toda2007, takamichi2015modulation, Tanaka2017}, frequency warping~\cite{tian2014correlation, Tian2015, Tian2016}, and recently deep neural networks (DNN)~\cite{Chen2014, Nakashika2013, mohammadi2014voice, dblstm_vc, ZHANG202031}. 
Other techniques, such as sparse representation~\cite{Wu2014, Sisman2018, 8688454}, and two-step adaptation~\cite{toda2006eigenvoice,kuhn2000rapid,tian2018average}, are studied to reduce the training data size. 
Non-parallel voice conversion techniques are certainly more attractive as parallel data is not easily available in practice. 
There has been prior work on finding the optimal segments from unpaired source and target training data, such as frame mapping~ \cite{erro2010inca,erro2007frame,qian2011frame}, clustering~\cite{abe1990voice,shikano1991speaker,turk2006robust} and generative adversarial network methods~\cite{cycle, cyclegan, cyclegan2, cyclegan3, stargan}. 

Speech carries speaker-independent linguistic information as well as speaker characteristics and speaking style. If we can disentangle linguistic features from speaker representation, we may synthesize the linguistic content from one voice to another. Variational autoencoder (VAE) ~\cite{hsu2016voice,hsu2017voice} represents a technique in this direction. It is built of an encoder-decoder neural network, where the encoder learns a latent space to represent the speaker-independent linguistic information, while the decoder reconstructs the target speech features by conditioning on a target speaker representation. AutoVC is another successful attempt, 
that is trained with a carefully designed bottleneck layer forcing the encoder to disentangle the speaker-independent features only with the self-reconstruction loss~\cite{autovc, autovc2}. It outperforms the VAE-based techniques without the need of linguistic supervision. The success of the autoencoder based framework is based on two assumptions: 1) the latent space only captures speaker-independent linguistic representation without a trace of the speaker information; 2) by conditioning on target speaker representation, a decoder is capable of generating the desired acoustic features for a target speaker. 
Unfortunately, the latent codes in autoencoder techniques are not trained to associate with linguistically motivated sound units. While autoencoder methods are effective in disentanglement, the lack of linguistic grounding could be a limitation for its performance.

The recent advances in deep learning approaches open up new possibilities beyond the traditional parallel and non-parallel voice conversion paradigms. One is able to transfer knowledge from external resources, such as automatic speech recognition (ASR) and text-to-speech (TTS) synthesis, to voice conversion systems.

Phonetic PosterioGram (PPG) draws a lot of attention in non-parallel voice conversion, that is an intermediate result from a speaker-independent ASR system, representing the posterior probability of phonetic classes of a speech frame~\cite{hazen2009query}. 
The PPG-based voice conversion techniques~\cite{Chinese,tian2018average, saito2018non} is an effective way to leverage  ASR knowledge, that is learnt from a large speech corpus. Despite their success, PPG techniques still suffer from some inherent limitations. For example, the quality of PPG highly depends on the ASR system that is neither optimized for speech synthesis nor voice conversion. 


Text-to-speech and voice conversion share a common goal to generate natural speech. However, TTS systems are usually trained with large speech corpora, while voice conversion systems face limited training data constraint. Therefore, transfer learning from TTS to voice conversion is naturally motivated in practice. Generally speaking, a neural TTS is trained to address two research problems, one is the ability to generate effective intermediate representations from the input text, another is the ability to align the attention
to the intermediate representations to bridge between encoder and decoder. In short, the former is referred to as feature mapping, the latter is referred to as alignment. There is a general belief that voice conversion can benefit from TTS in one way or another. For feature mapping, voice conversion can learn from TTS to produce phonetically motivated intermediate representations, which are speaker-independent; For alignment, voice conversion can benefit from the learned TTS for natural speech rendering and text-to-speech alignment. 

Park et al.~\cite{park2020cotatron} proposed a transcription-guided speech encoder as part of a sequence-to-sequence TTS model for any-to-many voice conversion, which requires both text and speech as input during run-time inference. 
On the other hand, Luong et al.~\cite{9004008} proposed to bootstrap a voice conversion system from a pre-trained speaker-adaptive TTS model, where both voice conversion and TTS share a common decoder. This method only handles feature mapping yet leaves the alignment task to an external system.
Zhang et al.~\cite{Zhang2019} proposed an architecture for joint training of voice conversion and text-to-speech. By taking text as an additional input, the voice conversion system improves voice quality during run-time inference.  However, it relies on large parallel training data. 
Huang et al.~\cite{huang2019voice} proposed a transformer architecture with TTS pre-training. The idea is to transfer knowledge from a pre-trained TTS model to a voice conversion model benefiting from large-scale, easily accessible TTS corpora. Though it attempts to handle alignment issues such as the conversion of articulation and prosody, this technique only works for parallel training data. 



Building on the success of the prior studies on speech disentanglement and TTS-VC transfer learning, in this paper, we study a novel transfer learning technique from TTS to voice conversion. We adopt Tacotron-2 as the TTS framework~\cite{tacotron2}. In a two-step training strategy, we first train a standard multi-speaker Tactron-2 on a large database. We then transfer the TTS knowledge to an encoder-decoder architecture for voice conversion. We hypothesize that, 1) the context vector generated by the text encoder in a TTS system is  speaker-independent representing the linguistic information;
2) the decoder in a TTS system, that constructs target acoustic features by conditioning on target speaker embedding, also works for voice conversion.

The main contributions of this paper include: 1) a novel transfer learning framework from TTS to voice conversion; 2) an integrated solution that benefits from transfer learning and speaker disentanglement; 3) an effective solution that deals with both feature mapping and alignment for voice conversion with non-parallel training data.

This paper is organized as follows. We introduce the related work that motivates our research in Section \ref{secre}. We formulate the proposed TTS-VC transfer learning framework in Section \ref{secpro}. We present the experiments and discuss the results in Section IV. Section \ref{secconc} concludes the study.

\begin{figure}
    \centering
    \includegraphics[width=.48\textwidth]{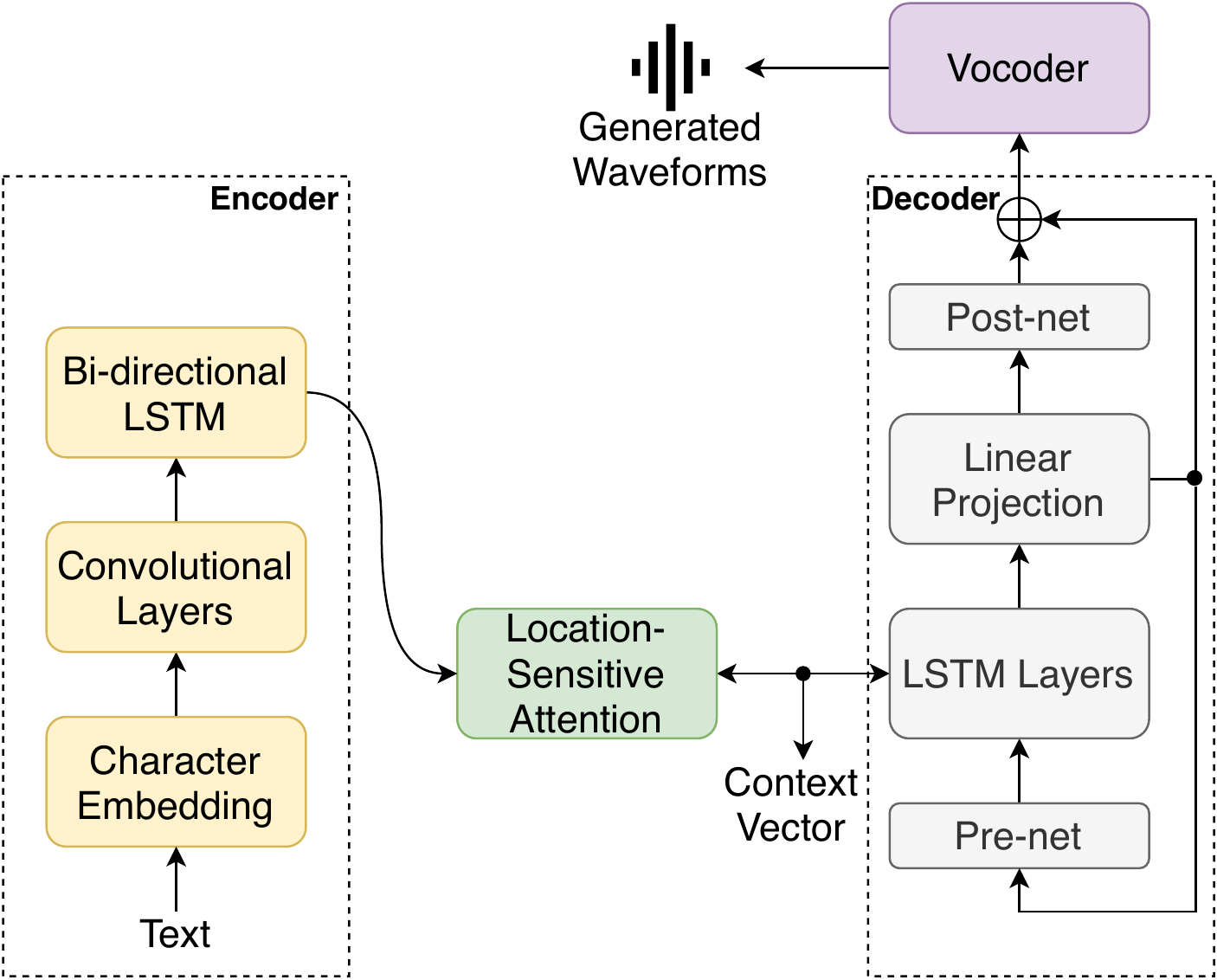}
    \caption{Block diagram of the Tacotron-2 system. A text-speech alignment can be obtained by feedings input text to encoder, and target spectral frames to the decoder in a teacher-forcing mode. This process produces a sequence of speaker-independent context vectors, that are frame-aligned with target mel-spectrogram, to serve as the supervision target in TTS-VC transfer learning.}
    \label{fig:tacotron}
\end{figure}

\section{Related Work}
\label{secre}
\subsection{Tacotron-2}

Tacotron is an end-to-end text-to-speech (TTS) system~\cite{wang2017tacotron} that has a sequence-to-sequence encoder-decoder architecture with attention mechanism~\cite{bahdanau2014neural}. Tacotron-2 is an updated version of Tacotron~\cite{tacotron2}, 
as illustrated in Fig. \ref{fig:tacotron}. It includes an encoder that maps an input text sequence to a fixed-dimensional state vector,  an attention-based decoder~\cite{NIPS2015_5847} that predicts a mel-spectrogram, and a neural vocoder~\cite{van2016wavenet} that reconstructs speech waveform from the predicted mel-spectrogram. For rapid turn-around, we use WaveRNN neural vocoder to generate speech waveform from mel-spectrogram in this paper~\cite{wavernn}.

The input to the encoder is a sequence of characters, which are first encoded into character embeddings. These embedded vectors are then processed by a stack of convolutional layers with batch normalization~\cite{bn}. The outputs of convolutional layers are taken by a bi-directional Long Short-Term Memory (LSTM) layer to produce text encoding.
\begin{figure}
    \centering
    \includegraphics[width=80mm]{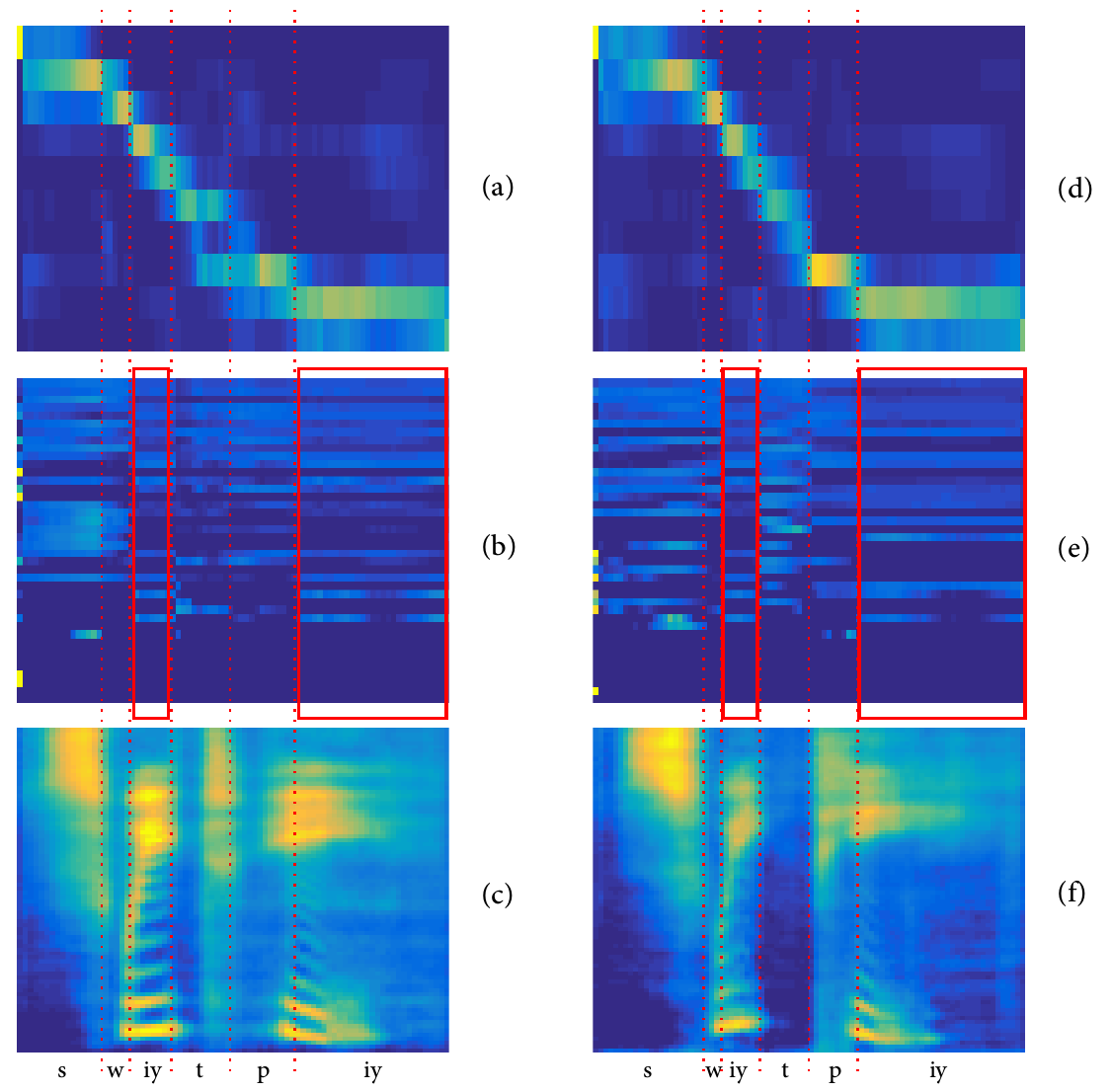}
    \caption{Illustration of an utterance \textit{`sweet pea'}  in (a), (b), and (c) for Speaker 1, and in (d), (e), and (f) for Speaker 2. (a) temporal alignment weight between text and mel-spectrogram,  (b) a stack of context vectors for the utterance (reduced to 40-dimension for ease of comparison), and (c) its mel-spectrogram with 80-dimensional spectral features.  The horizontal axis all plots represents time, while vertical axis of (b), (c), (e) and (f) displays the dimension of vectors.}
    \label{fig:context_mel}
\end{figure}

The decoder is an auto-regressive recurrent neural network (RNN) that predicts a mel-spectrogram from the text encoding. A location-sensitive attention mechanism is used to summarize the text encoding into a sequence of fixed-length context vectors. At each prediction time step, the previous predicted spectral frame and the current context vector are both used as the input to the decoder to predict the target spectral frame. All spectral frames form an output mel-spectrogram.


\subsection{Linguistic Representation}

In Tacotron-2 training, the text encoder converts an input sentence to a stack of fixed-dimension text encodings. Then the location-sensitive attention in Fig. \ref{fig:tacotron} learns to align the text encodings with a sequence of target spectral frames and generates the context vectors as a result. The context vectors are obtained by multiplying the alignment weights with text encodings. Fig. \ref{fig:context_mel} (a) and (d) show an example of the alignment between input text characters and mel-spectrogram learned by the attention mechanism. 

During training, the text-speech alignment can be obtained from the trained model in the teacher-forcing mode. In practice, we present the input text and the ground truth target mel-spectrogram to the encoder and decoder, respectively. As the decoder generates output spectral features in an autoregressive manner, we use the ground truth target mel-spectrogram in place of the predicted frames from previous time steps to guide the feature generation. Fig.~\ref{fig:context_mel} (b) and (e) illustrate a stack of generated context vectors, and Fig. \ref{fig:context_mel} (c) and (f) show their corresponding 80-dimensional mel-spectrogram. We note that the length of the context vector sequence is the same as that of the corresponding mel-spectrogram. 

As text encodings are derived to represent input text, that seeks to represent linguistic information. Context vectors are weighted sums over encoder time steps of the text encodings, that could be influenced by the mel-spectrogram, or acoustic features, via the decoder. However, such influence from acoustic features is minimum. It is generally considered~\cite{park2020cotatron} that context vectors represent speaker-independent linguistic features. This is especially true when we train a multi-speaker TTS using an encoder-decoder architecture, where the decoder is conditioned on speaker embedding \cite{multitts, 0shot}. In this case, the encoder-decoder architecture seeks to disentangle speaker from linguistic content.  

It is important that the context vectors are associated with input characters and optimized for speech synthesis task during the decoder training. In other words, they are linguistically motivated and serve as an ideal candidate of intermediate representation for encoder-decoder style voice conversion. Park et al. ~\cite{park2020cotatron} make use of context vectors from a TTS system as the linguistic features, that improve non-parallel voice conversion. In Fig.~\ref{fig:context_mel}, we use an utterance to compare the plot of a context vector sequence with its corresponding spectrogram for two speakers. We have three observations: a) the patterns of context vectors within the same phoneme (e.g. `s`, `w`, and `iy`) are rather stationary; b) the patterns for the phonemes are discrete and distinct from one another; c) the context vectors for the same phonemes are very close within the same speaker (see `iy` in the red boxes in Fig.~\ref{fig:context_mel} (b) or Fig.~\ref{fig:context_mel} (e))  and across speakers (see `iy` in the red boxes between Fig.~\ref{fig:context_mel} (b) and Fig.~\ref{fig:context_mel} (e)). All the observations point to the fact that the context vectors are the unique identifier of the phonemes, that are linguistically motivated. The context vectors don't reflect the spectral details of the actual acoustic rendering of the speech by individual speakers as illustrated in Fig.~\ref{fig:context_mel} (c) and (f). Our analysis corroborates the observations in ~\cite{park2020cotatron}.

In voice conversion, there have been previous studies on the use of PPG from automatic speech recognition (ASR) for linguistic representation of speech content~\cite{Chinese,tian2018average, saito2018non}. Context vector is a linguistic representation similar to PPG, but derived from text-to-speech synthesis (TTS) instead of ASR.

\subsection{Leveraging Knowledge from Speech Synthesis }

Traditionally, voice conversion operates at the signal level, while speech synthesis involves phonetic representation. Studies show that the use of linguistically informed features improves voice conversion. There have been studies to couple voice conversion with TTS training, that seeks to improve the training and run-time inference of voice conversion by adhering to linguistic content. However, such techniques usually require a large training corpus. 

Zhang et al.~\cite{zhang2019improving} proposed to improve the sequence-to-sequence model~\cite{zhang2019sequence} by using text supervision during training. A multi-task learning structure is designed which adds auxiliary classifiers to the middle layers of the sequence-to-sequence model to predict linguistic labels as a secondary task. The linguistic labels can be obtained with external alignment tools. With the linguistic label objective, the encoder and decoder are expected to generate meaningful intermediate representations which are linguistically informed. The text transcripts are only required during training. Zhang et al.~\cite{Zhang2019}, and Luong et al.~\cite{9004008} proposed joint training of TTS and VC by sharing a common decoder. Park et al.~\cite{park2020cotatron} proposed to use the context vectors in Tacotron system as speaker-independent linguistic representation to guide the voice conversion. 

Transfer learning is a technique to utilize knowledge from previously learned tasks and apply them to newer, related ones.  A typical transfer learning involves pre-training of a base model, reusing the pre-trained model as the starting point for a model on the second task of interest, and refining the second model on input-output pair data for the second task. In a TTS-VC transfer learning scenario, usually we have significantly more data for TTS, and we would like to generalize the learned knowledge from TTS training for VC, which has significantly less training data. Huang et al.~\cite{huang2019voice} proposed a technique to use a trained TTS decoder as the starting point of a VC decoder to train an encoder-decoder VC system. The study was focused on  conversion of a specific source-target pair with parallel training data. It doesn't aim to disentangle speaker and linguistic information. Nonetheless, it represents a successful attempt in TTS-VC transfer learning. 

All the studies suggest that voice conversion benefits from linguistically informed intermediate representations, and point to a direction for more studies on how voice conversion can benefit from TTS systems without involving large training data. TTS-VC transfer learning becomes a natural choice that will be the focus of this paper.  

\subsection{Speaker Disentanglement and Voice Cloning}

Speaker disentanglement and voice cloning with autoencoder~\cite{hsu2016voice,hsu2017voice} represents one of the successful techniques in voice conversion with non-parallel training data, where the encoder learns to disentangle speaker representation from speaker-independent linguistic representation, and the decoder reconstructs target speech with linguistic features, conditioning on target speaker representation. A speaker encoder is usually used to generate such speaker representation, e.g. speaker embeddings. Successful examples include i-vector~\cite{Dehak2011}, x-vector, and d-vector \cite{variani2014deep}. 

With speaker disentanglement, the decoder is able to reconstruct speech for any target speakers unseen during training, that we call voice cloning. Voice cloning has also been a successful technique in speech synthesis that takes text as input and generates voice of unseen speakers, when given a few speech samples \cite{voiceclone, luong2020nautilus}. As the idea of voice cloning with speaker embeddings are proven effective in both TTS and VC, a common network architecture certainly facilitates the TTS-VC transfer learning.

\section{TTS-VC Transfer Learning}
\label{secpro}

Voice conversion is typically a research problem with scarce training data, however, deep learning techniques are typically data driven, that rely on big data. This is actually the strength of deep learning in voice conversion. 
Deep learning opens up many possibilities to benefit from abundantly available training data, so that the voice conversion task can focus more on learning the mapping of speaker characteristics. For example, we wouldn't like the voice conversion task to infer low level detail during speech reconstruction, a neural vocoder can learn from a large database to do so~\cite{wavenet-vae2019}. We wouldn't like the voice conversion task to learn how to represent an entire phonetic system of a spoken language either, a general purpose acoustic model from a speech recognition or synthesis system can learn from a large database to do a better job. 

The formulation of transfer learning aims to achieve just that. By leveraging the large database, we free up the conversion network from using its
capacity to represent low level detail and general information, but instead, to focus on the high level semantics necessary for speaker identity conversion.

\begin{figure*}
    \centering
    \includegraphics[width=.95\textwidth]{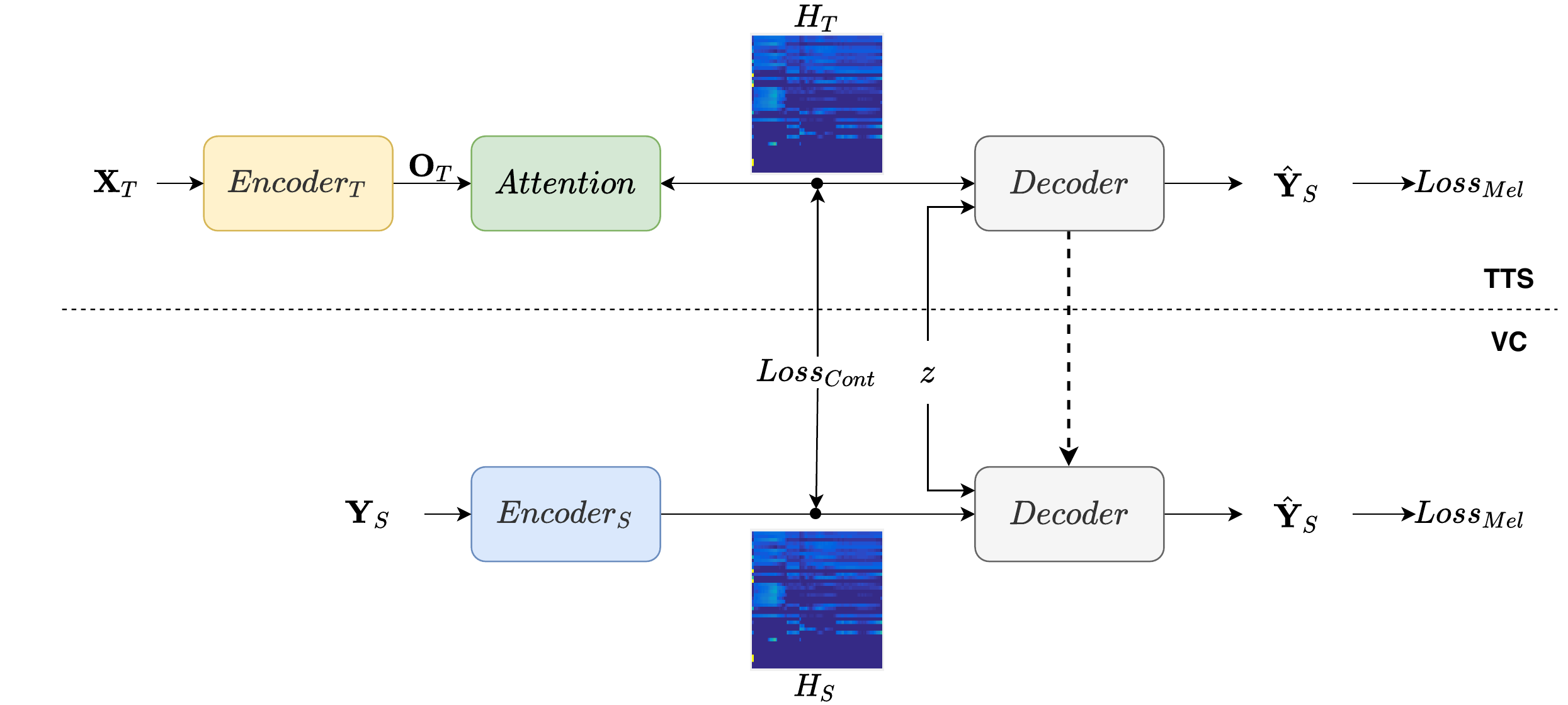}
    \caption{Diagram of the proposed TTS-VC transfer learning architecture. The upper panel is a Tacotron TTS pipeline, and the lower panel is a voice conversion pipeline.  $\mathbf{X}_{T}$ denotes input text, $\mathbf{Y}_{S}$ and $\hat{\mathbf{Y}}_{S}$ are target mel-spectrogram and the mel-spectrogram generated by the pipelines; $\mathbf{O}_{T}$ denotes text encoding,  $\mathbf{H}_{T}$ denotes the context vectors from TTS pipeline, $\mathbf{H}_{S}$ denotes the context vectors equivalents from VC pipeline; $z$ denotes speaker embedding.}
    \label{fig:proposed}
\end{figure*}

In this section, we will describe the proposed TTS-VC transfer learning architecture and a two-step training scheme. Fig. \ref{fig:proposed} illustrates the architecture of our proposed model and the loss function adopted during model training.

\subsection{Pre-training of the Multi-Speaker TTS model}

An encoder-decoder TTS model offers two useful properties: 1) The TTS decoder is trained to produce linguistic features from the text that is assumed speaker independent;  2) A multi-speaker TTS decoder provides a way to combine speaker-independent linguistic features and speaker embedding to produce speech in target voice. From voice conversion point of view, we would like to disentangle speaker-independent linguistic features from source speech and re-compose them with target speaker embedding to generate speech in target voice. The linguistic features and the decoder mechanism are the knowledge that voice conversion would like to learn from.

Tacotron-2 model was firstly studied for single speaker TTS \cite{tacotron2}. To train a multi-speaker Tacotron model, we consider the use of the speaker embedding and where to apply the speaker embedding.
We adopt the same speaker verification network~\cite{g2e} as in~\cite{multitts} to generate a fixed-dimensional speaker embedding vector. In~\cite{multitts}, speaker embedding is applied on encoder output before the attention mechanism, hence, the resulting context vectors are speaker dependent. In this paper, we would like to generate context vectors that are sufficiently speaker independent. Therefore, we propose to incorporate speaker embeddings only after the attention mechanism as the controlling input to the TTS decoder. In this way, the encoder-decoder TTS architecture serves as a disentangling mechanism, which uses context vectors to represent speaker-independent linguistic features, and speaker embedding to represent the speaker's voice identity.

As shown in Fig. \ref{fig:proposed}, the text encoder transform a sequence of text characters $\mathbf{X}_{T} = \{\mathbf{x}_{T}^{1}, \mathbf{x}_{T}^{2}, ... \mathbf{x}_{T}^{M}\}$ to a sequence of fixed-dimension embedding vectors:
\begin{equation}
\begin{aligned}
    \mathbf{O}_{T} &= Encoder_{T}(\mathbf{X}_{T}) \\
    &= \{\mathbf{o}_{T}^{1}, \mathbf{o}_{T}^{2}, ..., \mathbf{o}_{T}^{M}\}
\end{aligned}
\end{equation}
where $M$ denotes the length of the text sequence.

With the attention mechanism, we can obtain an alignment between the text embeddings and the mel-spectrogram features, that is described by a weight matrix $\mathbf{W}$. The context vectors $\mathbf{H}_{T}$ can be obtained by applying the weight matrix on the text embeddings:
\begin{equation}
\begin{aligned}
    \mathbf{H}_{T} &= \mathbf{W} \times \mathbf{O}_{T} \\
    &= \{\mathbf{h}_{T}^{1}, \mathbf{h}_{T}^{2}, ..., \mathbf{h}_{T}^{N}\}
\end{aligned}
\end{equation}
where $N$ denotes the length of the mel-spectrum frames. 
During training, we feed all mel-spectrum frames to the pre-net in a teacher-forcing mode, where $N$ is the length of the training utterance. At run-time inference, $N$ is predicted by the decoder.

The decoder takes the concatenation of context vectors $\mathbf{H}_{T}$ and speaker embedding $z$ to generate the mel-spectrum features, $\hat{\mathbf{Y}}_{S}$:
\begin{equation}
\begin{aligned}
    \hat{\mathbf{Y}}_{S} &= Decoder(concat(\mathbf{H}_{T}, z)) \\
    &= \{\hat{\mathbf{y}}_{S}^{1}, \hat{\mathbf{y}}_{S}^{2}, ..., \hat{\mathbf{y}}_{S}^{N}\}
\end{aligned}
\end{equation}

The loss function is defined as the mean square error (MSE) between the ground truth mel-spectrogram $\mathbf{Y}_{S}$ and the predicted one $\hat{\mathbf{Y}}_{S}$:
\begin{equation}
\label{lossmel}
\begin{aligned}
    Loss_{Mel} &= MSE(\mathbf{Y}_{S}, \hat{\mathbf{Y}}_{S}) \\
    &= \frac{1}{N}\sum_{n=1}^{N}||\mathbf{y}_{S}^{n} - \hat{\mathbf{y}}_{S}^{n}||^{2}
\end{aligned}
\end{equation}

With a trained Tacotron-2 model, to get the context vector of the training data, we feed input text to the encoder and ground truth mel-spectrum to the decoder in a teacher-forcing mode. By doing this, we obtain a sequence of context vectors that has the same length as the sequence of mel-spectrum intended for voice conversion training.

\subsection{Transfer Learning from TTS to VC}
We propose an encoder-decoder voice conversion framework similar to those in ~\cite{hsu2016voice, Chou2018, huang2019voice} in terms of the system architecture. The VC encoder seeks to generate speaker-independent linguistic features from input spectral features, while the VC decoder reconstructs the mel-spectrum features from the linguistic features, conditioning on a speaker code. Studies show that voice conversion benefits from explicit phonetic modeling that ensure adherence to linguistic content during conversion~\cite{park2020cotatron, 9004008}. 

The question is how to establish the correspondence between input mel-spectrum features and the speaker-independent linguistic features. The PPG-based voice conversion techniques~\cite{Chinese,tian2018average, saito2018non} require an external ASR system to work along side during training and inference. The autoencoder-style voice conversion frameworks learn the latent codes in an unsupervised manner, therefore, they are either speaker independent or phonetically motivated. Others rely on an explicit temporal alignment process~\cite{S2S}.

Unlike the prior work, we propose to use the linguistic features, i.e. context vectors, from a trained TTS model, that are phonetically motivated, to serve as the supervision target of the VC latent code during training. By doing this, VC benefits from TTS model in many ways, as shown in Fig. 3. First, the trained TTS provides a temporal alignment between input text $\mathbf{X}_{T}$ and context vectors $\mathbf{H}_{T}$, the latter is frame-aligned with mel-spectrum features $\mathbf{Y}_{S}$; Second, the context vectors $\mathbf{H}_{T}$ represent speaker-independent linguistic features of input text, that are suitable to serve as the supervision targets of $\mathbf{H}_{S}$ for the voice conversion encoder; Third, the TTS decoder can be used as the initialization of VC decoder; Fourth, the same TTS training dataset can be used for VC training without the need of additional dataset. The voice conversion encoder can be described as follows,
\begin{equation}
\begin{aligned}
    \mathbf{H}_{S} &= Encoder_{S}(\mathbf{Y}_{S}) \\
    &= \{\mathbf{h}_{S}^{1}, \mathbf{h}_{S}^{2}, ..., \mathbf{h}_{S}^{N}\}
\end{aligned}
\end{equation}
A loss function is introduced to minimize the distance between a VC latent code $\mathbf{H}_{S}$ and context vector $\mathbf{H}_{T}$:
\begin{equation}
\label{losscont}
\begin{aligned}
    Loss_{Cont} &= MSE(\mathbf{H}_{S}, \mathbf{H}_{T}) \\
    &= \frac{1}{N}\sum_{n=1}^{N}||\mathbf{h}_{S}^{n} - \mathbf{h}_{T}^{n}||^{2}
\end{aligned}
\end{equation}

In this work, the VC decoder is similar to the TTS decoder functionally, that takes the concatenation of the linguistic features $\mathbf{H}_{S}$ and the speaker embedding $z$ to generate the mel-spectrum features $\hat{\mathbf{Y}}_{S}$:
\begin{equation}
\begin{aligned}
    \hat{\mathbf{Y}}_{S} &= Decoder(concat(\mathbf{H}_{S}, z)) \\
    &= \{\hat{\mathbf{y}}_{S}^{1}, \hat{\mathbf{y}}_{S}^{2}, ..., \hat{\mathbf{y}}_{S}^{N}\}
\end{aligned}
\end{equation}

The TTS decoder is trained with the TTS pipeline, that takes $\mathbf{H}_{T}$ as input. We propose to use the TTS decoder as the VC decoder and the VC encoder is trained to produced TTS latent codes. However, there could be a potential mismatch between the TTS decoder and $\mathbf{H}_{S}$. To minimize such mismatch, we use the TTS decoder as initialization, and refine the VC decoder through an adaptation process, with the same loss function as Eq. (\ref{lossmel}).

To summarize, the speech encoder and decoder are trained with the joint loss function as formulated in Eq. (\ref{lossmel}) and (\ref{losscont}).

\begin{equation}
    \begin{aligned}
        Loss_{Joint} &= Loss_{Cont} + Loss_{Mel} \\
        &= \frac{1}{N}\sum_{n=1}^{N}(||\mathbf{h}_{S}^{n} - \mathbf{h}_{T}^{n}||^{2} + ||\mathbf{y}_{S}^{n} - \hat{\mathbf{y}}_{S}^{n}||^{2})
    \end{aligned}
\end{equation}
The training procedure seeks to learn to disentangle linguistic information from speaker information and to optimize the representations for speech generation.

During the transfer learning, we use the same data as those for TTS pre-training. The difference is that the text transcript is no longer required in VC training.  No additional speech data is required either.

\subsection{Zero-shot Run-time Inference}
Once the TTS-VC transfer learning is completed, the voice conversion pipeline is able to perform voice conversion independently without involving the attention mechanism of TTS. 
During inference, both the source and the target speaker might be unseen speakers, that is referred to as any-to-any voice conversion. The proposed framework is able to perform such any-to-any voice conversion without further system training, that is also called zero-shot run-time inference.

To convert an utterance from source to target, we only need a speech sample, e.g. one utterance, from the target speaker. We use the speech sample to obtain a speaker embedding $z_{t}$ from a speaker verification network. The run-time inference stage can be formulated as:
\begin{equation}
    \hat{\mathbf{Y}} = Decoder(concat(Encoder_{s}(\mathbf{Y}), z_t))
\end{equation}
where $\mathbf{Y}$ denotes the source mel-spectrogram and $\hat{\mathbf{Y}}$ denotes the converted target mel-spectrogram.

\subsection{Spectral and Prosodic Mapping}
Traditionally voice conversion is performed by an analysis-mapping-reconstruction pipeline, where source speech is first decomposed into vocoding parameters such as spectral features, $F_{0}$ and aperiodicity indicators.  Individual vocoding parameters are then mapped from source to target by respective conversion models. Statistical model, regression model, and deep learning model are commonly used.
Recently, end-to-end TTS shows that it is possible to predict both spectral and prosodic features from a latent representation by a decoder~\cite{prosodytaco}. This suggests that voice conversion can also be done in the same way if we are able to characterize the input speech with similar latent representation. In this paper, with TTS-VC transfer learning, we adopt the strategy in end-to-end TTS to model spectrum and prosody simultaneously. We consider $\mathbf{H}_{T}$ speaker-independent and $F_{0}$ agnostic. Hence, $\mathbf{H}_{S}$, which is learnt under the supervision of $\mathbf{H}_{T}$, is also expected to be $F_{0}$ agnostic. In this way, the decoder is capable to model $F_{0}$ without being influenced by $F_{0}$ from the source speaker. The decoder is followed by WaveRNN vocoder to reconstruct a speech signal. 

\subsection{Comparison with other VC systems}

The proposed TTS-VC transfer learning framework, denoted as TTL-VC for short, represents a new way to leverage TTS knowledge. It employs a simple architecture to couple with TTS for knowledge transfer and an independent voice conversion pipeline for inference. To draw a clear distinction between TTL-VC and other prior work, we provide a comprehensive comparison in terms of data requirement, system configuration, and application scenarios in Table \ref{table:algo-comparison}. The four systems in Table \ref{table:algo-comparison} are discussed in Section II as the prior work. They represent the recent progress in TTS-assisted voice conversion. From Table \ref{table:algo-comparison}, we note that TTL-VC is unique in many ways. With transfer learning, TTL-VC doesn't rely on TTS encoder during run-time inference, it is trained solely on the TTS training data without the need of additional training data; With disentangled latent codes, TTL-VC is able to perform any-to-any voice conversion without involving model re-training or adaptation for unseen speakers. 

The four systems in Table~\ref{table:algo-comparison} do not employ transfer learning, furthermore, they have different requirements about training data. Therefore, a direct comparison of their performance is not meaningful. Instead, we benchmark the proposed TTL-VC with three competing systems, as summarized in Table \ref{table:ex-algo-comparison}, that share the same decoding strategy, as formulated in Eq. (9), and are trained on the same dataset for a fair comparison.

\begin{table*}[!htb]
\centering
\caption{The properties of voice conversion systems that leverage TTS knowledge. TTL-VC transfer learning supports flexible voice conversion task (from one unseen speaker to another) at run-time, while keeping minimum data requirement (non-parallel training data, no adaptation for unseen speakers, only speech is required during inference).}
\begin{tabular}{c|c|c|c|c|c|c}
\hline\hline
    & \begin{tabular}[c]{@{}c@{}} \bf TTS-VC\\ \bf transfer learning\end{tabular} & \begin{tabular}[c]{@{}c@{}} \bf source-target\\ \bf training data \end{tabular} & \begin{tabular}[c]{@{}c@{}} \bf VC inference\\ \bf input\end{tabular} &   \begin{tabular}[c]{@{}c@{}} \bf TTS involved\\ \bf during inference \end{tabular} & \begin{tabular}[c]{@{}c@{}} \bf additional training\\ \bf data for VC\end{tabular} & \bf VC application\\ \hline
Park et al. ~\cite{park2020cotatron}  & no & non-parallel & text and speech & encoder and decoder & yes & any-to-many\\ \hline
Luong et al.~\cite{9004008} & no & non-parallel  & speech & decoder & yes & any-to-one \\ \hline
Zhang et al.~\cite{Zhang2019} & no & parallel &  text and speech  & decoder & yes & many-to-one \\ \hline
Huang et al.~\cite{huang2019voice} & no & parallel & speech & decoder & yes & one-to-one\\ \hline
TTL-VC & yes & non-parallel  & speech  & decoder & no & any-to-any \\ \hline\hline
 
\end{tabular}
\label{table:algo-comparison}
\end{table*}

\section{Experiments}
\subsection{Experimental Setup and Model Architecture}

The three competing baselines in Table \ref{table:ex-algo-comparison} include a multi-speaker TTS model, a PPG-VC model and an AutoVC model. They represent the state-of-the-art voice conversion performance. They also share the same idea, that is to use speaker-independent linguistic features as the latent codes, in order to support any-to-any voice conversion.   

The speaker verification network used to generate speaker embedding \cite{g2e} is a 3-layer LSTM network with 768 hidden units followed by a linear projection layer with a size of 256. The resulting d-vector serves as speaker embedding $z_t$ for all systems in Table \ref{table:ex-algo-comparison}. Next, we briefly describe the experimental setup in this comparative study.

\begin{table*}[!htb]
\centering
\caption{Performance benchmarking between TTL-VC and three competitive baselines. All systems are trained on the same multi-speaker non-parallel dataset, and share a similar decoding architecture that takes latent linguistic features, conditioning on speaker embedding, to generate target speech. Notes: MS-TTS and TTL-VC use text transcripts of speech data during training; MS-TTS takes text as input, while others take speech as input. }
\begin{tabular}{c|c|c|c|c|c}
\hline\hline
                    &  \bf latent code & \bf latent code training & \bf training data & \bf decoder  & \bf inference input  \\ \hline
MS-TTS~\cite{tacotron2}  &   phonetic context vectors & supervised & text-speech pair & autoregression w/ attention & text  \\ \hline
PPG-VC~\cite{Chinese}        &  phonetic posteriogram  &  supervised & PPG-decoded speech & autoregression w/o attention & speech \\ \hline
AutoVC~\cite{autovc}                & content code &  unsupervised &  speech only & framewise mapping w/o attention & speech \\ \hline
{TTL-VC} & phonetic context vectors & supervised & text-speech pair  & autoregression w/o attention  & speech \\ \hline\hline
 
\end{tabular}
\label{table:ex-algo-comparison}
\end{table*}

\subsubsection{TTS-VC Transfer Learning (TTL-VC)}

We employ a two-step training scheme for TTL-VC system. 

First, a multi-speaker TTS (MS-TTS) is trained as a teacher model, that follows Tacotron-2 architecture~\cite{tacotron2} as illustrated in the upper panel of Fig. 3. The encoder converts input text to a sequence of 512-dimensional character embeddings. These embeddings pass through 3 1-dimensional convolutional layers, each containing 512 filters with 5 kernel size, followed by batch normalization \cite{bn} and GELUs activation~\cite{gelu}. The convolutional layer output is taken by a single bi-directional LSTM layer with 512 hidden units. The pre-net of the decoder is a stack of 2 fully-connected layers with 256 hidden units followed by GELUs activation. The LSTM layers in the decoder contain 2 uni-directional LSTM layers with 1024 hidden units. The linear projection layer outputs the 80-dimensional mel-spectrogram. The post-net contains 5 1-dimensional convolutional layers each containing 512 filters with 5 kernel size, followed by batch normalization and \textit{tanh} activation.

Second, the voice conversion pipeline in TTL-VC takes speech as input and generates speech as output, that is illustrated in the lower panel of Fig. 3, where input and output speech is represented as 80-dimensional mel-spectrum features. We conduct the transfer learning as discussed in Section III.B. The VC encoder is trained to generate context vectors that are similar to those in TTS pipeline; the VC decoder is trained to take the context vectors and speaker embedding $z_t$ to recompose the target speech. Both encoder and decoder adopt the same architecture of the MS-TTS model. No attention mechanism is involved in voice conversion pipeline as it only performs a framewise mapping, thus no temporal alignment is required.

\subsubsection{Multi-speaker TTS model (MS-TTS)}

The MS-TTS model is the teacher model in TTL-VC. We note that MS-TTS is not a voice conversion system, it takes the text and speaker embedding as input, and generates mel-spectrum features as output. We adopt it as a baseline for two reasons.

First, while the MS-TTS system shares the same decoder architecture as TTL-VC, its context vectors $\mathbf{H}_T$ are produced only from the text without influence from any source speaker. We expect that the synthesized speech to be highly similar to that of target speaker; Second, by comparing the prosodic patterns between MS-TTS and TTL-VC, we would like to observe whether the attention mechanism in MS-TTS has an advantage over the framewise mapping in TTL-VC.

\subsubsection{PPG-VC model}
An ASR system is trained with the Kaldi toolkit~\cite{povey2011kaldi}. 
We first process the input speech into a sequence of 40-dimensional MFCC  features with 12.5 ms frame shift. The ASR system then converts the speech feature sequence into a phonetic posteriogram (PPG) sequence, where a PPG frame represents the probability distribution over 132 phonetic classes. In PPG-VC, a pre-trained ASR serves as the encoder. Only the VC decoder is involved in the training. The VC decoder is similar to the TTS decoder in Tacotron-2~\cite{tacotron2}. The decoder LSTM layer takes PPG frames as input, and generate 80-dimensional mel-spectrum as output.


We adopt the PPG-VC model as one of the baseline models because it shares a similar encoder-decoder architecture as TTL-VC. Furthermore, PPG-based VC systems represent state-of-the-art performance in recent voice conversion challenges (VCC)~\cite{vcc2018}. 

\subsubsection{AutoVC model}
If we consider TTL-VC as an extension to the autoencoder (AE) based techniques, AutoVC serves as a competitive reference model. AutoVC, a successful AE-based implementation, is known for its impressive performance in speaker disentanglement~\cite{autovc}. With the same training data, the difference between TTL-VC and AutoVC mainly lies in the adoption of transfer learning. TTL-VC employs linguistically motivated context vectors while AutoVC obtains latent codes in an unsupervised manner. Simply speaking, the lower panel of Fig. 3 resembles the AutoVC workflow except that TTL-VC introduces $Loss_{Cont}$ during the encoder-decoder pipeline training process.  We implement AutoVC as a baseline to observe the benefit of TTS knowledge transfer.

For a fair comparison between TTL-VC and AutoVC~\cite{autovc}, our AutoVC configuration follows that of the TTL-VC network.  The dimension of the latent code and the up/down sampling factor are set to 32, the AutoVC decoder performs frame-wise mapping without auto-regression. In this way, the decoder only contains LSTM layers, linear projection and post-net.  Both TTL-VC and AutoVC take 80-dimensional mel-spectrum as input and generate mel-spectrum features as output.

\subsection{Database and Feature Extraction}
All systems in Table II and WaveRNN vocoder are trained on the same dataset from LibriTTS database \cite{libritts}. We use the \textit{train-clean-360} subset of the database that contains 191.29 hours of speech data from a total of 904 speakers, that consist of 430 female and 474 male speakers. The audio files are recorded at 24kHz sampling rate.

For system evaluation, we use speech data from the VCC2018~\cite{vcc2018}. We use the evaluation subset of the dataset that contains 8 source speakers (4 female speakers and 4 male speakers) and 4 target speakers (2 female speakers and 2 male speakers). Each speaker provides 35 utterances. Each source-target group, namely Female-Female, Female-Male, Male-Male, and Male-Female, has 4 speakers as the sources and 2 speakers as the targets, consisting of a total of 280 conversion pairs. The audio files are recorded at 22.05kHz sampling rate.

\begin{figure}
    \centering
    \includegraphics[width=85mm]{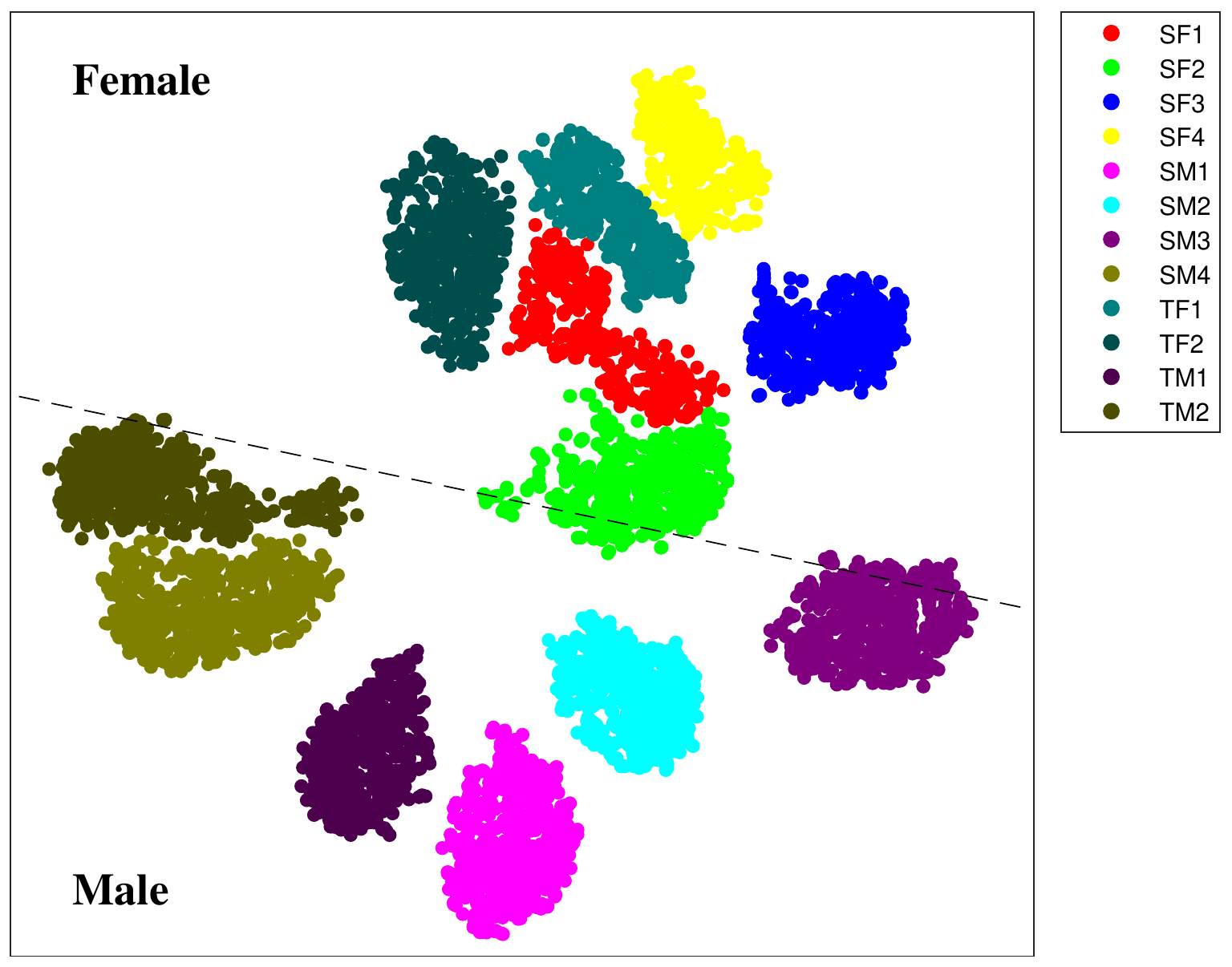}
    \caption{Visualization of the speaker embedding clusters using t-SNE for 12 speakers (SF1-4, SM1-4, TF1, TF2, TM1, and TM2) in VCC 2018 dataset.}
    \label{fig:embedding}
\end{figure}

We train a speaker verification network as speaker encoder on AISHELL2 \cite{du2018aishell} corpus. AISHELL2 contains 1,000 hours of speech data from 1,991 speakers, including 845 male speakers and 1,146 female speakers. We obtain an equal error rate (EER) of 2.31\% for speech samples, each having 3 seconds on average, on a test set of 63.8 hours that consists of 200 unseen speakers. Using the speaker encoder, we derive a speaker embedding $z_t$ from 5 seconds of speech sample for all 12 speakers in VCC 2018 database, and visualize them in Fig.~\ref{fig:embedding} using t-distributed stochastic neighbor embedding (t-SNE) algorithm \cite{tsne}. We observe a clear clustering of speech samples by speakers.

All speech data is resampled to 16kHz for PPG extraction. The acoustic features are extracted with 12.5ms frame shift and 50ms frame length.
The ASR model contains 4 bidirectional gated recurrent unit (GRU) layers with 512 hidden units in each layer.
Followed by a softmax layer, the 256-dimensional probability output is taken as PPG features.

\subsection{Results and Discussion}
\subsubsection{Objective Evaluation}
We evaluate the systems in terms of mel-cepstrum distortion (MCD) and root mean square errors of $F_{0}$ (RMSE)  between converted and reference speech utterances. MCD is defined as
\begin{equation} \label{eq:3}
MCD[dB] = 10/\mathrm{ln}10\sqrt{2\sum_{d=1}^{D}(\hat{Y}_d-Y_d)^2},
\end{equation}
where $D$ is the mel-cepstral coefficients (MCCs) feature dimension, $\hat{Y}_d$ and $Y_d$ are the $d^{\mathrm{th}}$ coefficients of the converted and original MCCs, respectively.
A lower MCD value accounts for a lower distortion \cite{mcd}.

The $F_{0}$ RMSE is defined as
\begin{equation} \label{eq:4}
RMSE[Hz] = \sqrt{\frac{1}{N}\sum_{i=1}^{N}(\hat{F}_{0_i}-{F}_{0_i})^2},
\end{equation}
where $N$ is the number of frames, $\hat{F}_{0_i}$ and ${F}_{0_i}$ are the corresponding $F_{0}$ values at the $i^{th}$ frame of the converted and reference speech, respectively.

\begin{figure*}
\centering
\subfloat[Source]{\includegraphics[width=70mm]{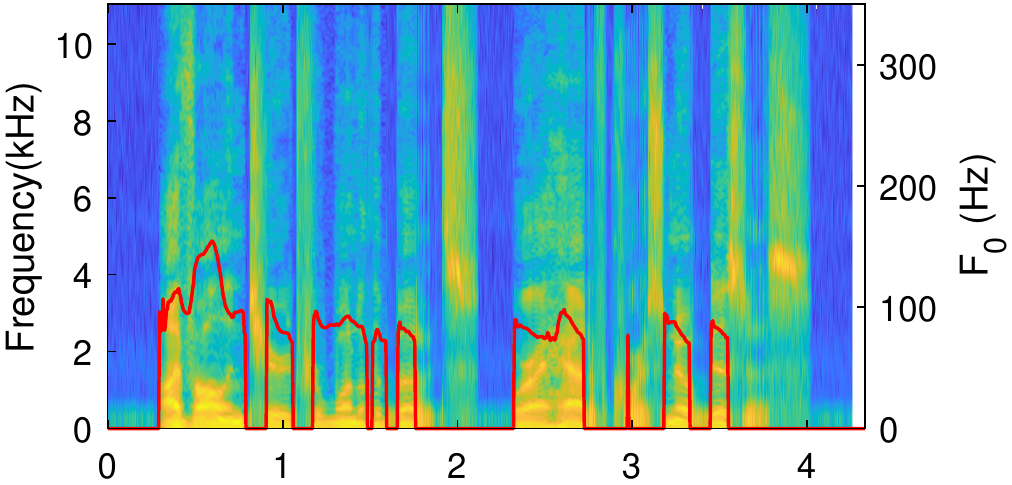}%
\label{fig:src}}
\hfil
\subfloat[Target]{\includegraphics[width=70mm]{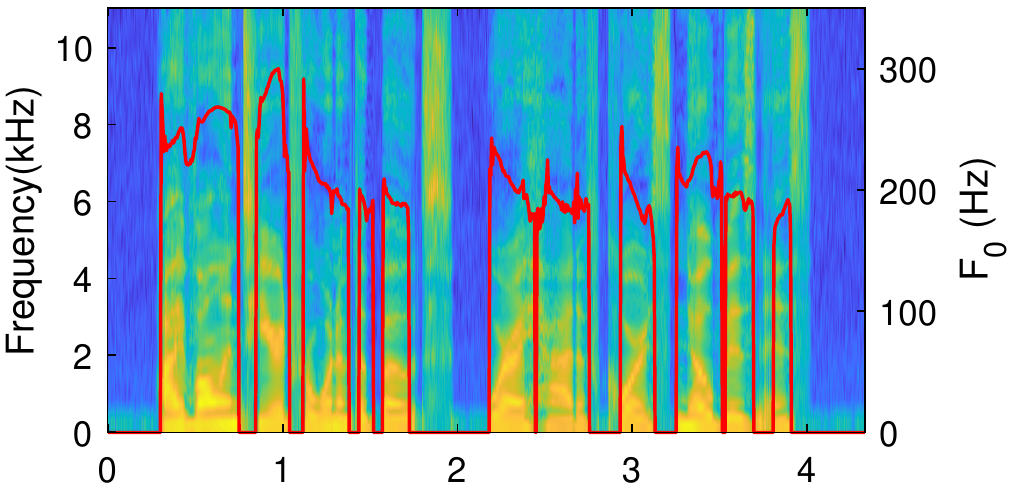}%
\label{fig:trg}}
\hfil
\subfloat[AutoVC]{\includegraphics[width=70mm]{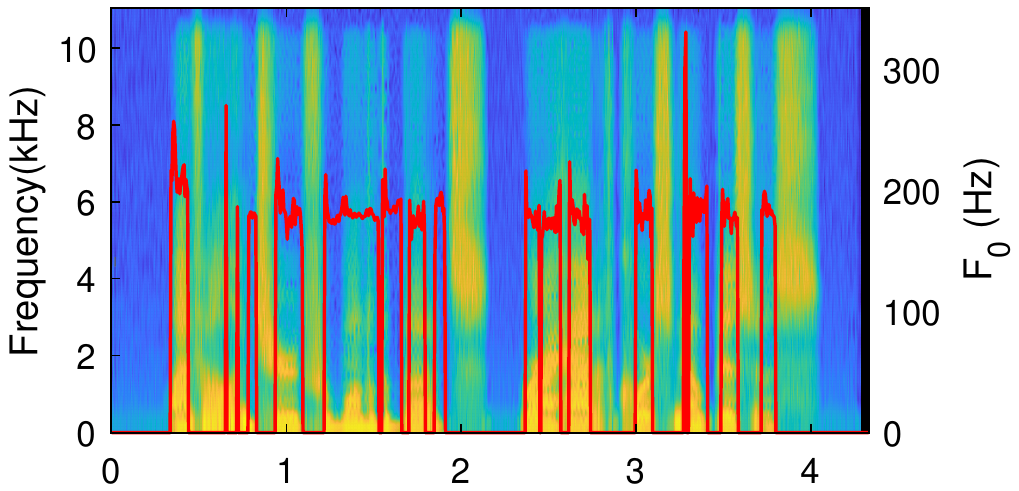}%
\label{fig:tts}}
\hfil
\subfloat[PPG-VC]{\includegraphics[width=70mm]{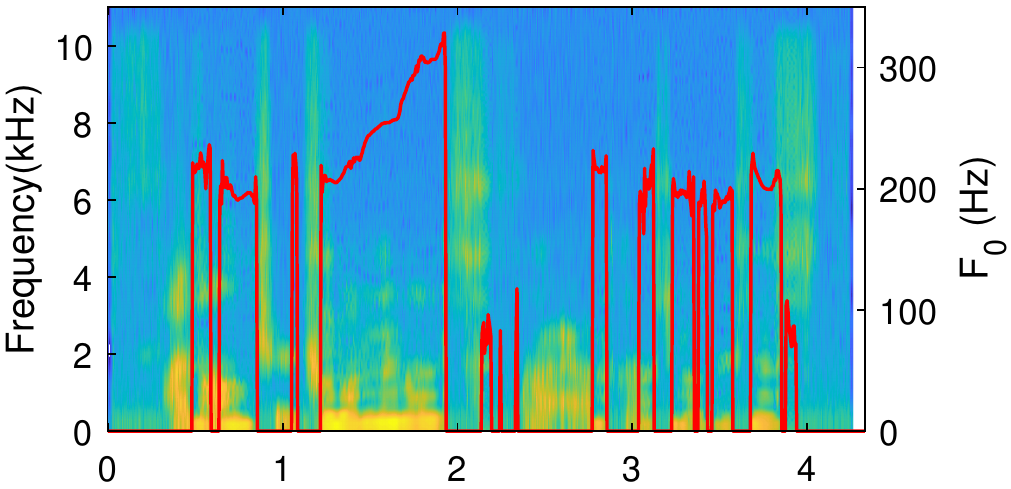}%
\label{fig:ppgvc}}
\hfil
\subfloat[MS-TTS]{\includegraphics[width=70mm]{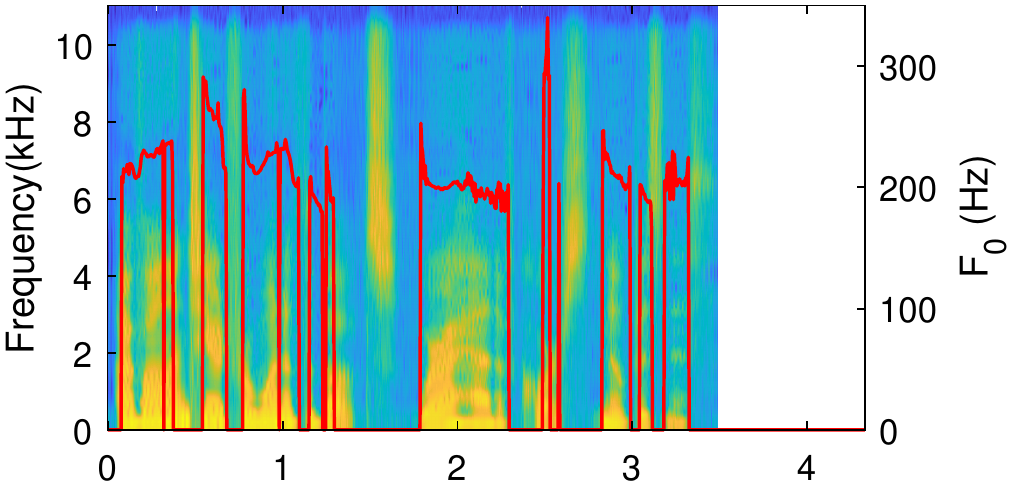}%
\label{fig:vaevc}}
\hfil
\subfloat[TTL-VC]{\includegraphics[width=70mm]{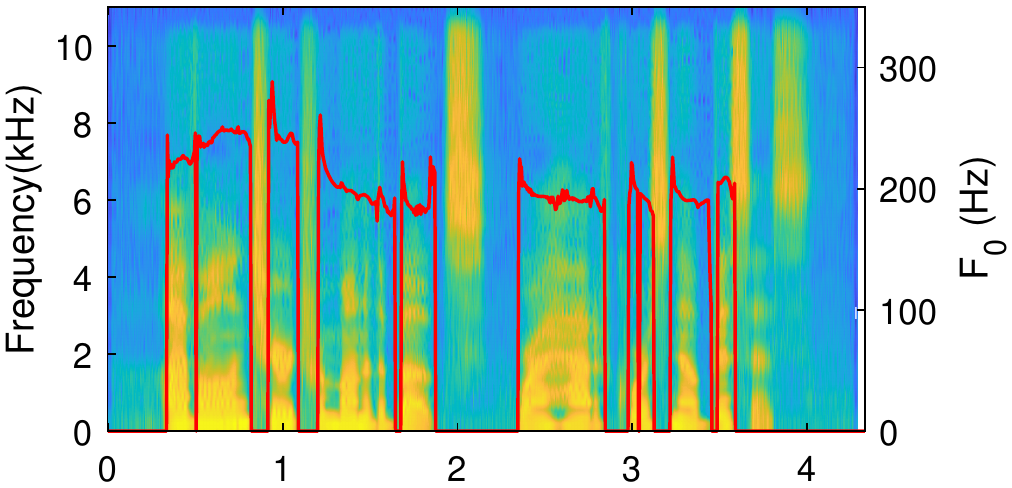}%
\label{fig:proposevc}}

\caption{The comparison of spectrogram and $F_{0}$ for a male-female voice conversion example `\textit{I volunteered for one of the boats, where I had, of course, no business}' using four models. 
Horizontal axis (x-axis) displays time in second, and vertical axis (y-axis) represents spectral frequency and $F_{0}$ frequency respectively.   } 
\label{fig:mel_f0}
\end{figure*}

As mel-spectrogram features are adopted as acoustic features, $F_{0}$ and MCCs are not readily available in the converted acoustic features. We extract $F_{0}$ and 25-dimensional MCCs using WORLD vocoder \cite{world} from the reconstructed waveform for evaluation purpose. To account for the temporal difference, dynamic time warping is performed between the converted utterance and the target reference to compute MCD and $F_{0}$ RMSE, where $F_{0}$ RMSE is calculated only on the voiced frames in the reference utterances. 
\begin{table}
\centering
\caption{Average MCD and $F_{0}$ RMSE of between the converted sample and the target reference for \textit{AutoVC}, \textit{PPG-VC} and \textit{TTL-VC} models. \textit{Source} denotes the distortion directly between source and target reference, which could be the worst situation as no conversion has taken place. }

\begin{tabular}{c|c|c|c}
\hline\hline
Source-Target                  & Model     & MCD (dB)      & $F_{0}$ RMSE (Hz)  \\ \hline
\multirow{4}{*}{Female-Female} & Source    & 8.49          & 50.67 \\ \cline{2-4}
                              & AutoVC    & 4.03          & 51.53  \\ \cline{2-4} 
                              & PPG-VC    & 4.16         & 59.39  \\ \cline{2-4} 
                              & TTL-VC    & \textbf{3.77} & \textbf{36.49}  \\ \hline
\multirow{4}{*}{Female-Male}   & Source    & 8.67          & 84.52 \\ \cline{2-4}
                              & AutoVC    & 3.46          & 61.03  \\ \cline{2-4} 
                              & PPG-VC    & 3.47          & 39.93  \\ \cline{2-4} 
                              & TTL-VC    & \textbf{3.31} & \textbf{17.37}  \\ \hline
\multirow{4}{*}{Male-Male}     & Source    & 9.62          & 34.96 \\ \cline{2-4}
                              & AutoVC    & 3.54          & 35.26  \\ \cline{2-4} 
                              & PPG-VC    & 3.55          & 42.64  \\ \cline{2-4} 
                              & TTL-VC    & \textbf{3.36} & \textbf{17.58}  \\ \hline
\multirow{4}{*}{Male-Female}   & Source    & 10.38         & 81.66 \\ \cline{2-4}
                              & AutoVC    & 4.19          & 58.79  \\ \cline{2-4} 
                              & PPG-VC    & 4.25          & 60.61  \\ \cline{2-4} 
                              & TTL-VC    & \textbf{3.96} & \textbf{35.99}  \\ \hline
{Text-Female}   & MS-TTS    & 4.32          & 52.06  \\ \hline
{Text-Male}     & MS-TTS    & 3.85          & 37.47  \\ \hline\hline
\end{tabular}
\label{table:mcd}
\end{table}

TABLE \ref{table:mcd} summarizes the MCD and RMSE evaluation as an average over the 280 conversion pairs (4 source speakers $\times$ 2 target speakers $\times$ 35 utterances) for each source-target gender group, and 70 utterances (2 target speakers $\times$ 35 utterances) for MS-TTS speech synthesis.  It is observed that  TTL-VC outperforms all other voice conversion models consistently for both MCD and $F_{0}$ RMSE. We note that female speakers have a higher $F_{0}$ variance than male speakers, that is reflected in the $F_{0}$ RMSE of MS-TTS samples, and generated speech for female target speakers. For both male and female target speakers, TTL-VC shows a clear advantage in prosodic mapping. We are glad to see that TTL-VC significantly outperforms TTS systems, which suggests that TTL-VC not only delivers the exact linguistic content but also with improved prosody over TTS output. 

By comparing TTL-VC and MS-TTS, we note that the difference lies in the encoder. The former takes speech as input, while the latter takes text as input. The results suggest that source speech is more informative than text input in providing speaker-independent prosodic patterns for target speech generation. This could be explained by the fact that the prosodic pattern of a sentence is a modulation between speaker-independent components, e.g. prosodic phrasing, intonation, and speaker-dependent components, e.g. accent, pitch level. Source speech provides essential speaker-independent components for the reconstruction of pitch contour, while text input doesn't.


To visualize the effect, Fig. \ref{fig:mel_f0} takes a male-female example to compare the spectrogram and $F_{0}$ of natural speech and generated speech from various models. We have three observations, 1) The duration of the TTS synthesized speech is predicted by an attention mechanism, it differs from that of either source or target natural speech. PPG-VC, AutoVC and TTL-VC generate speech that has the same duration as the source because they perform the framewise mapping. 2) {The $F_{0}$ prosodic patterns of both PPG-VC and AutoVC are closer to the source, with a $F_0$ RMSE of 52.75 and 70.48, than to the target, with a $F_0$ RMSE of 56.56 and 93.37 respectively. This suggests that both PPG and autoencoder latent codes are influenced} by the source speaker. 3) The $F_{0}$ prosodic pattern of TTL-VC is closer to the target ($F_0$ RMSE = 18.26) than to the source ($F_0$ RMSE = 122.29). This suggests that the context vectors are rather speaker independent, that allow the decoder to recompose speech well for target speaker using target speaker embedding. The observations in Fig. \ref{fig:mel_f0} are consistent with the performance statistics in Table \ref{table:mcd}.

\begin{figure*}
\centering
\subfloat[]{\includegraphics[width=53mm]{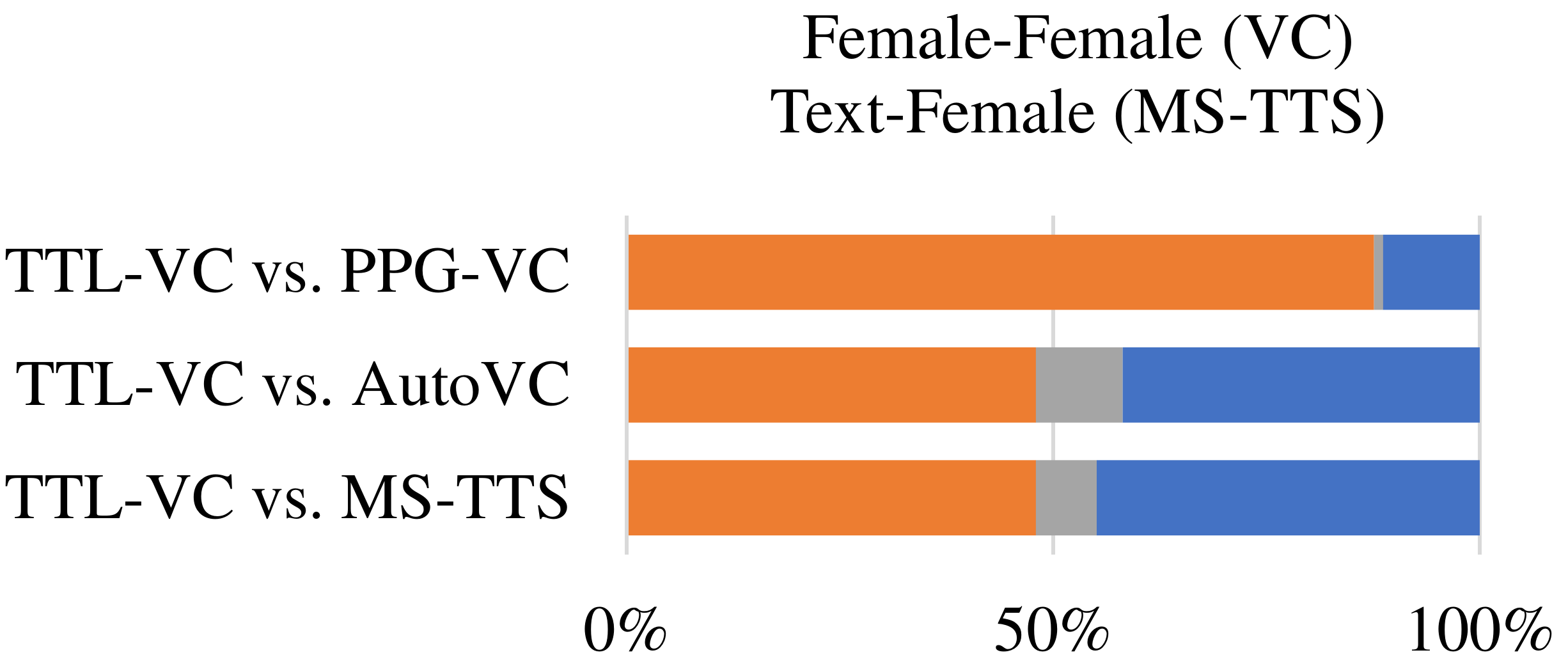}%
\label{fig:ab_f2f}}
\hfil
\subfloat[]{\includegraphics[width=30mm]{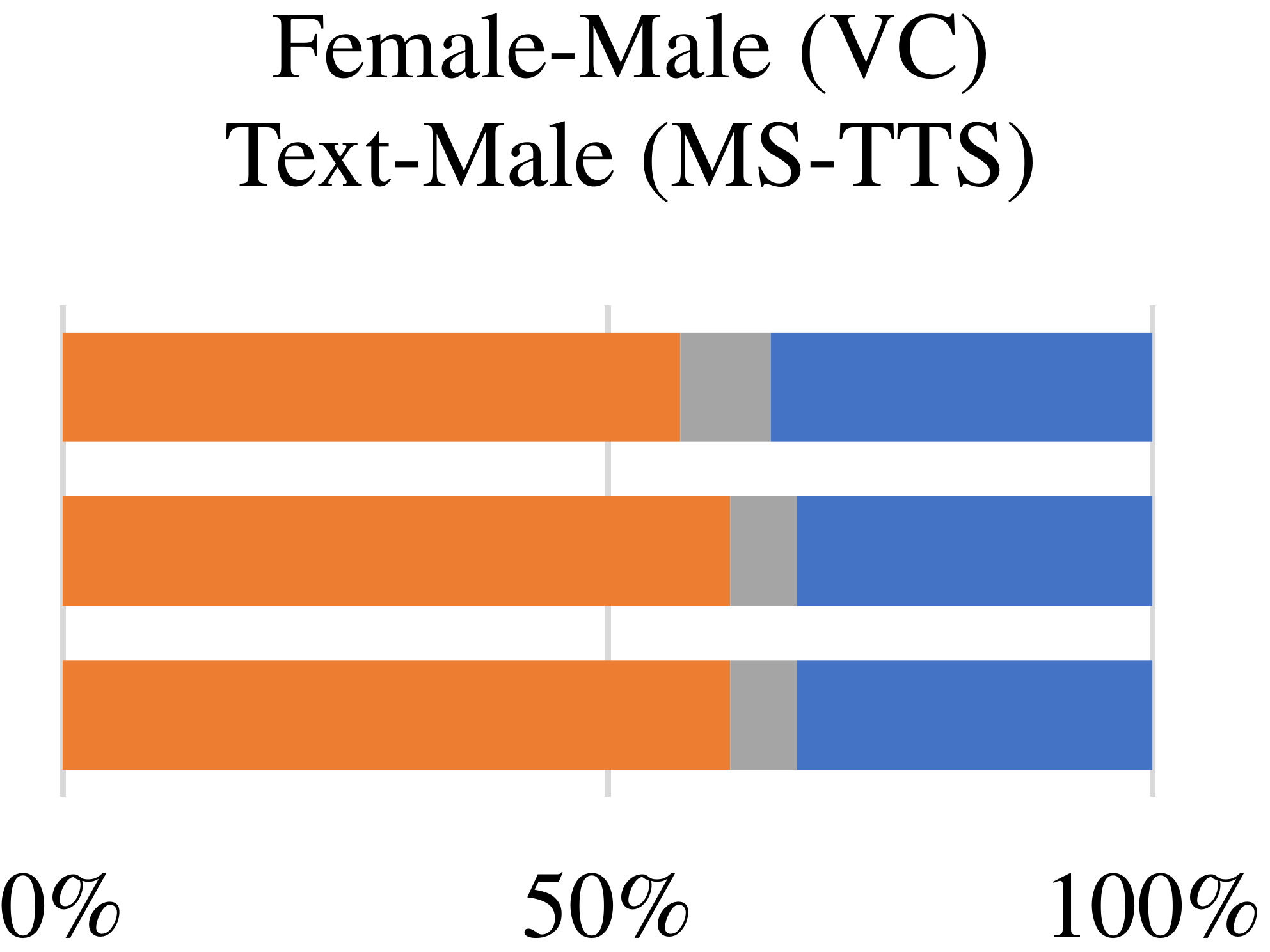}%
\label{fig:ab_f2m}}
\hfil
\subfloat[]{\includegraphics[width=30mm]{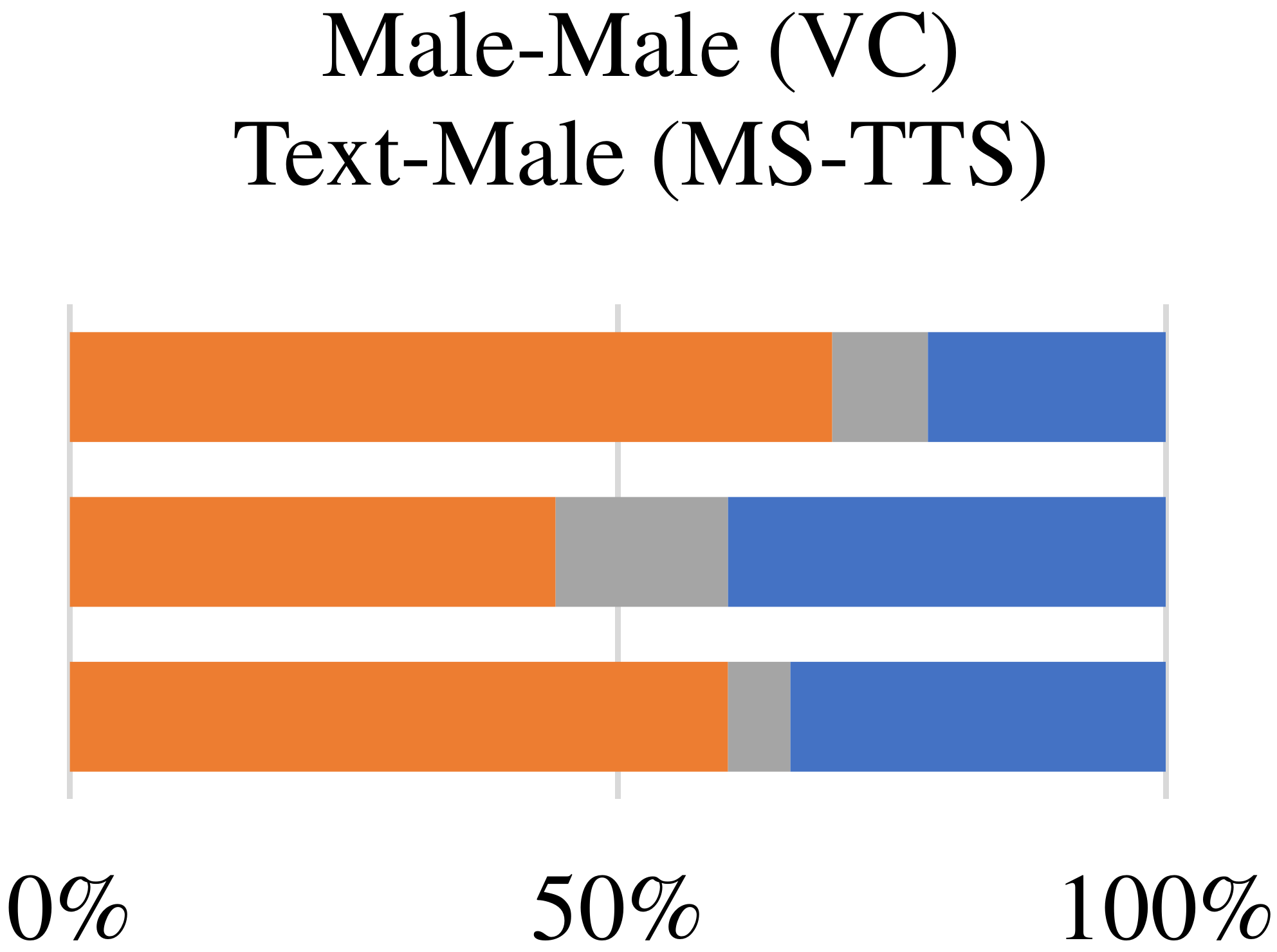}%
\label{fig:ab_m2m}}
\hfil
\subfloat[]{\includegraphics[width=48mm]{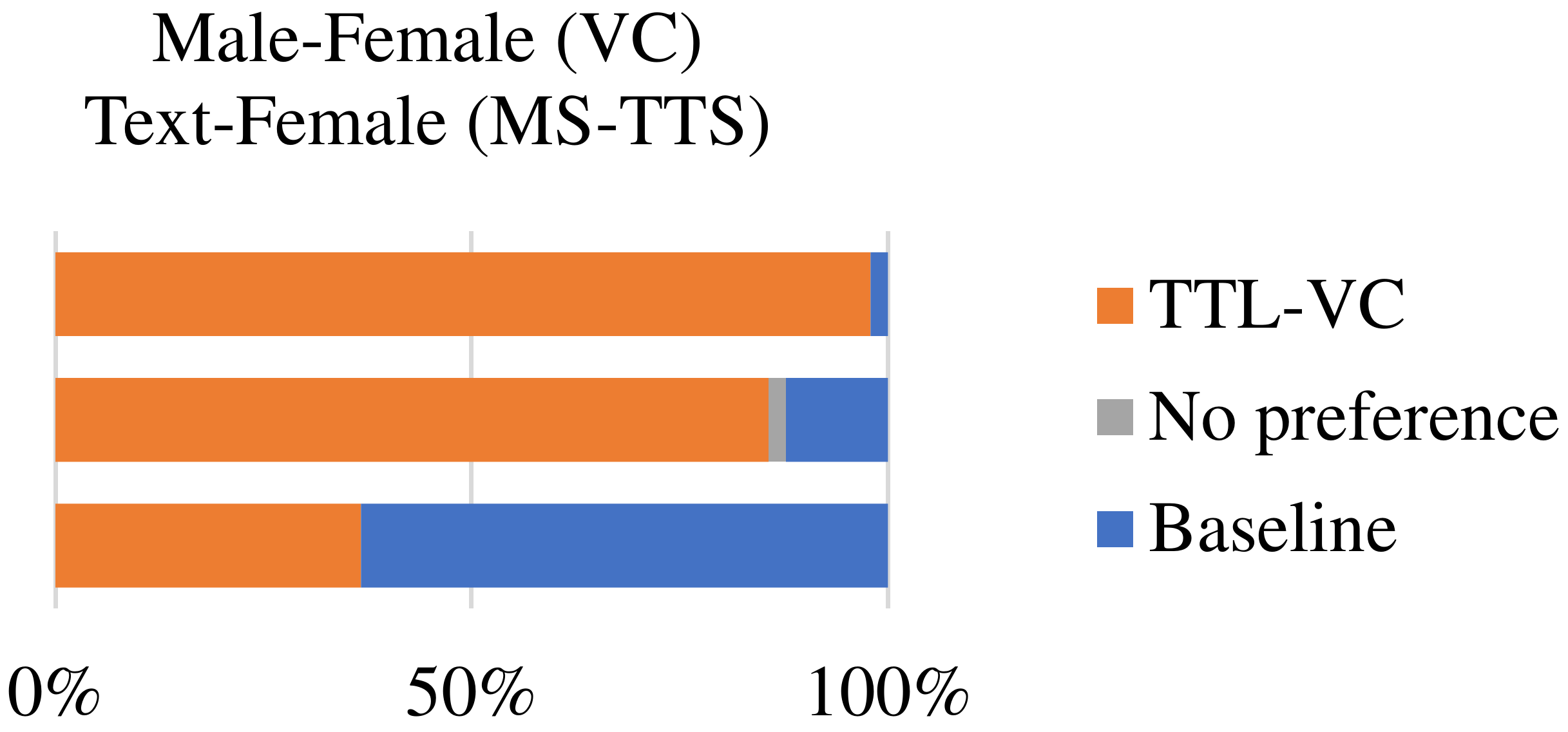}%
\label{fig:ab_m2f}}
\caption{AB test between TTL-VC and other models for four source-target groups.}
\label{fig:ab}
\end{figure*}

\begin{figure*}
\centering
\subfloat[]{\includegraphics[width=53mm]{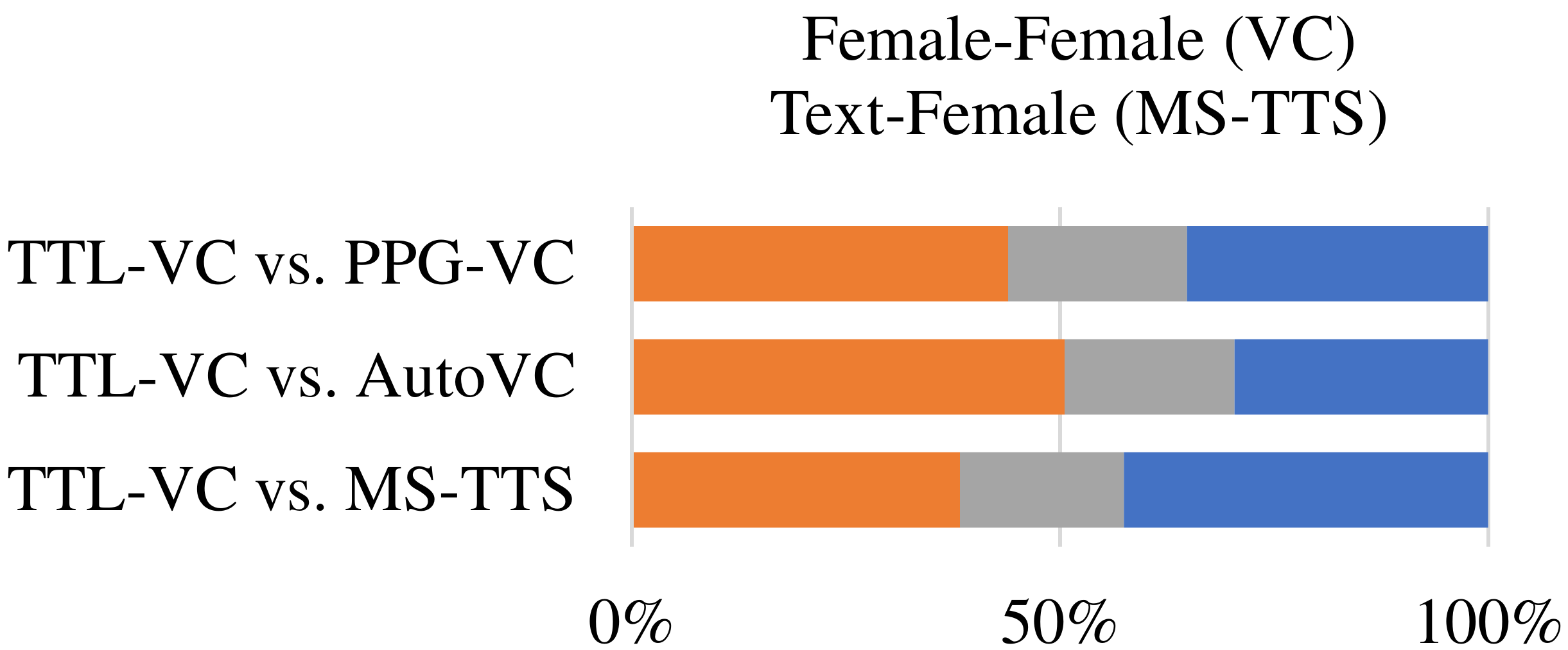}%
\label{fig:abx_f2f}}
\hfil
\subfloat[]{\includegraphics[width=30mm]{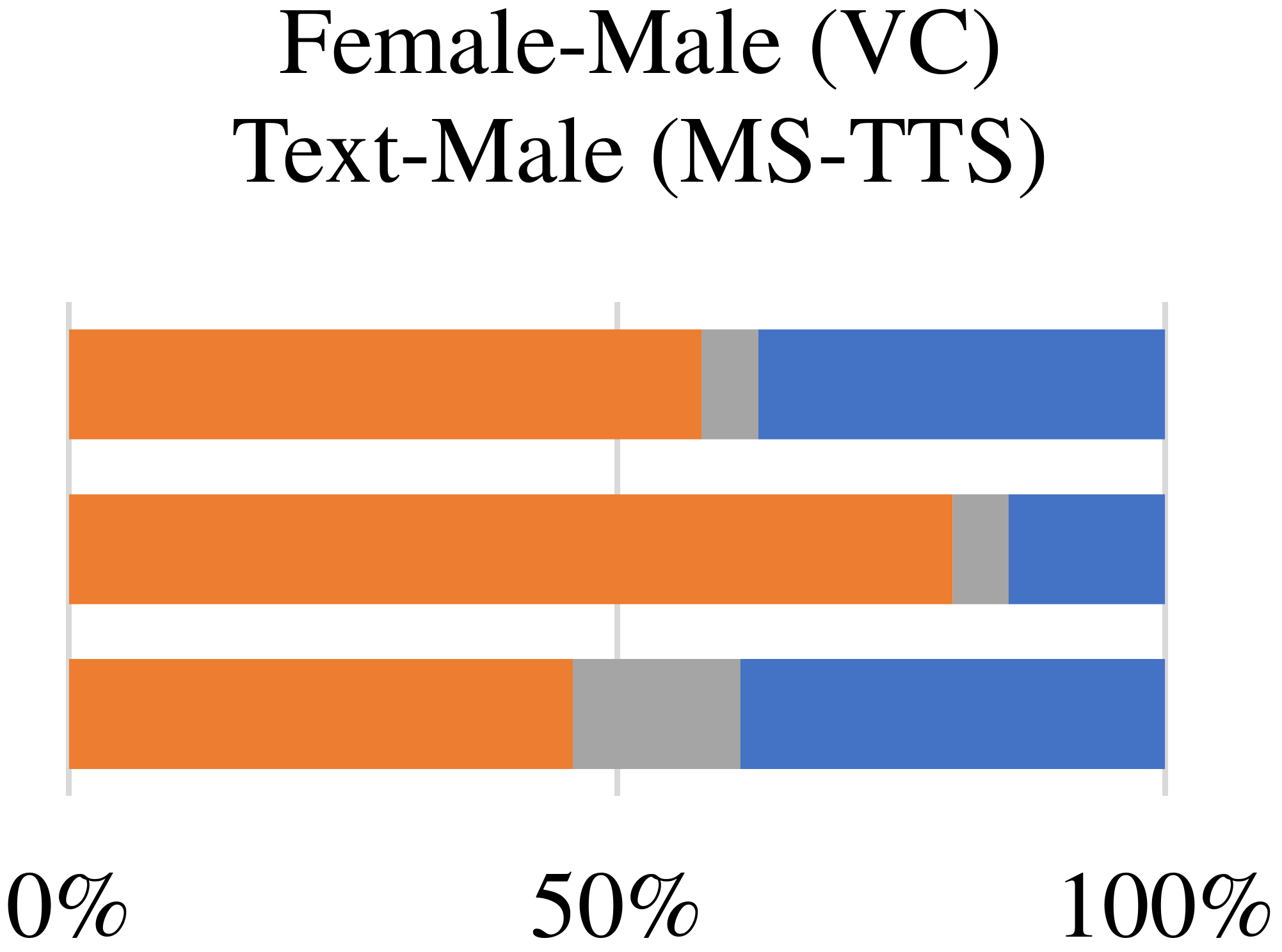}%
\label{fig:abx_f2m}}
\hfil
\subfloat[]{\includegraphics[width=30mm]{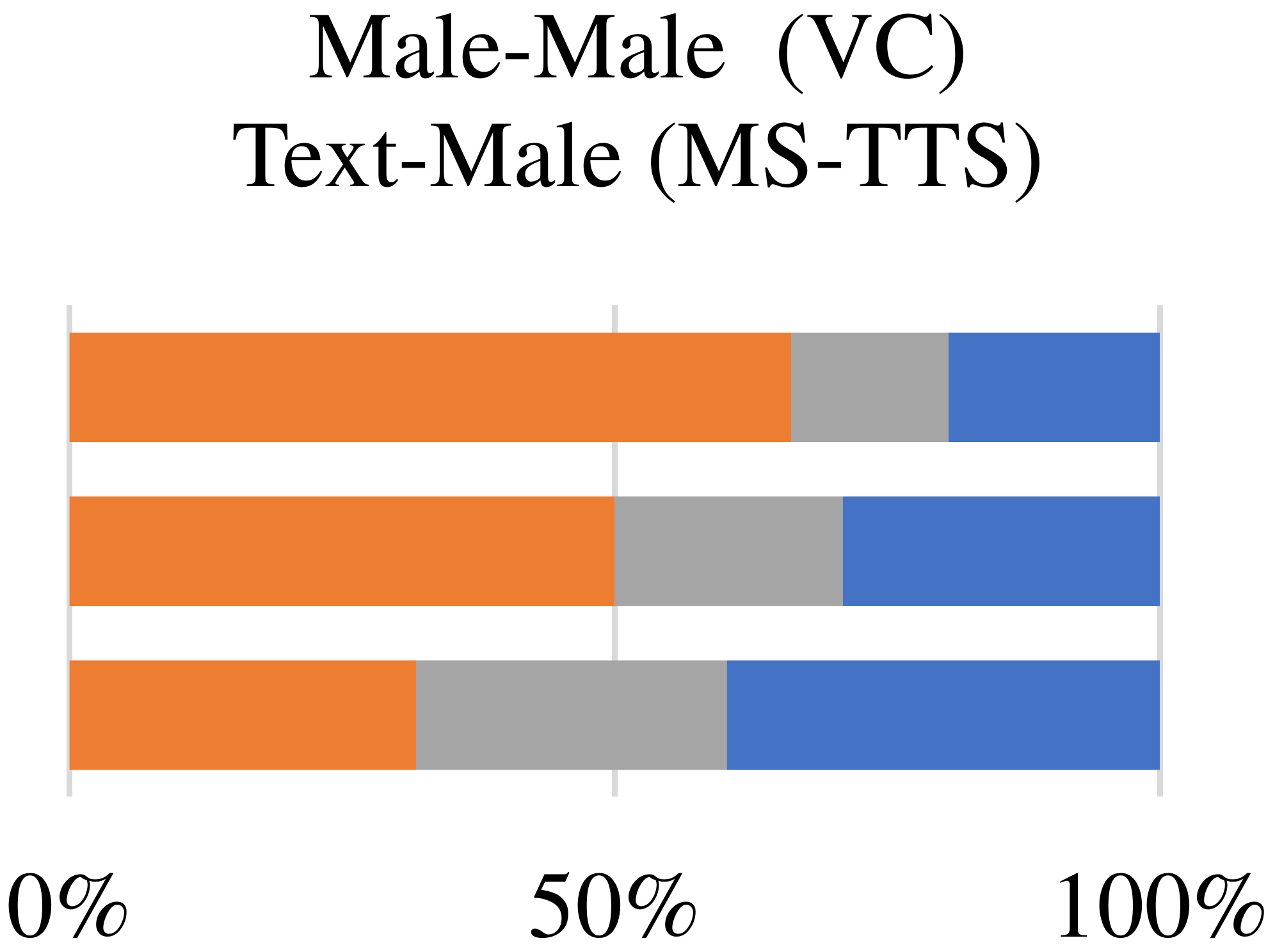}%
\label{fig:abx_m2m}}
\hfil
\subfloat[]{\includegraphics[width=48mm]{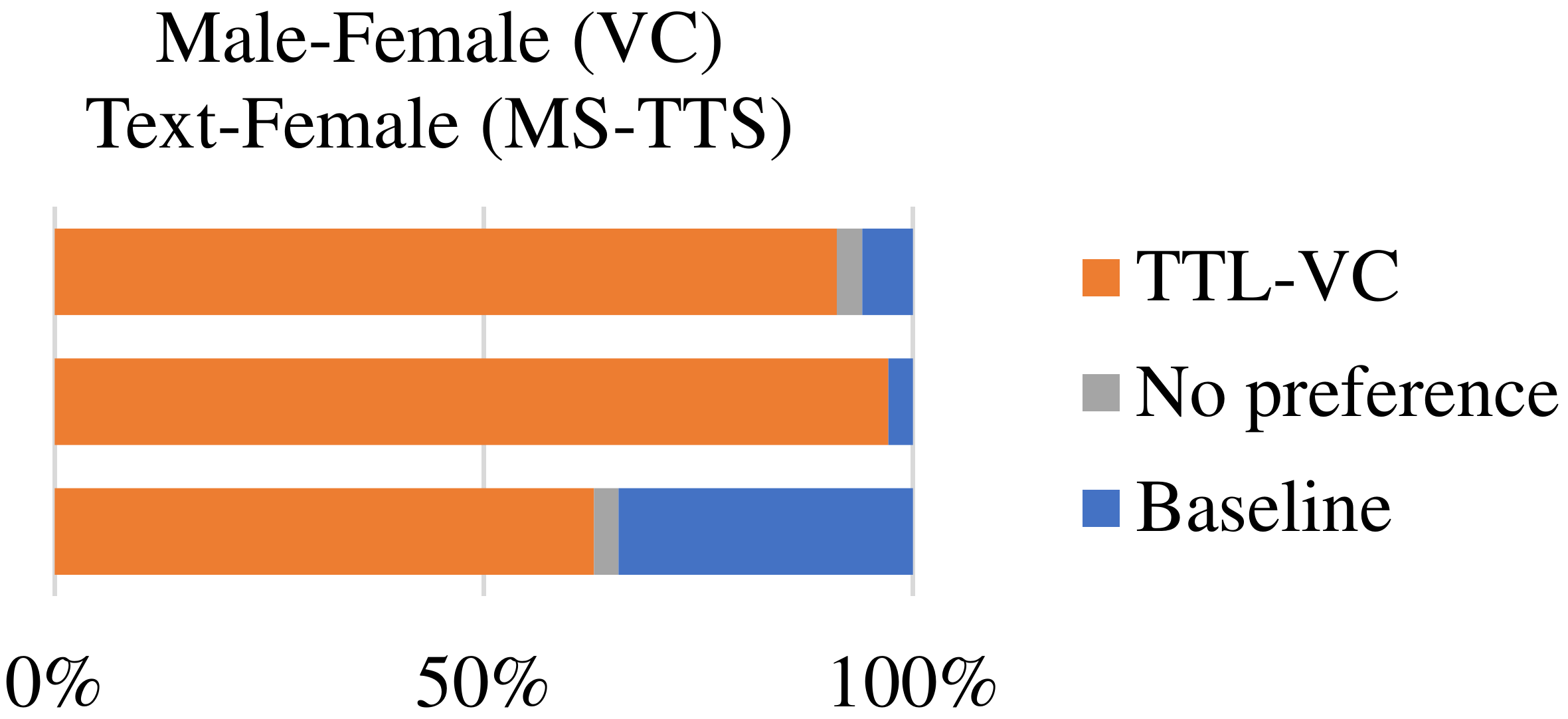}%
\label{fig:abx_m2f}}
\caption{XAB test between TTL-VC and other models for four source-target groups.}
\label{fig:abx}
\end{figure*}

\begin{table}
\centering
\caption{Best-worst scaling test results for the converted samples by source-target pair and their aggregate. The Best\% or Worst\% over four models sum to 100\%.}
\begin{tabular}{c|c|c|c}
\toprule[2pt]
Source-Target                & Model       & Best\%                  & Worst\%                      \\ \hline\hline
\multirow{3}{*}{Female-Female} & AutoVC & 26.47               & 11.03 \\ \cline{2-4} 
                               & PPG-VC & 5.52                & 77.57 \\ \cline{2-4} 
                               & TTL-VC & 40.07               & 5.52  \\ \hline
Text-Female                    & MS-TTS & 27.94               & 5.88  \\ \hline\hline 
\multirow{3}{*}{Female-Male}   & AutoVC & 7.57                & 57.37 \\ \cline{2-4} 
                               & PPG-VC & 25.90               & 29.88 \\ \cline{2-4}
                               & TTL-VC & 40.63               & 2.79  \\ \hline
Text-Male                      & MS-TTS & 25.90               & 9.96  \\ \hline\hline 
\multirow{3}{*}{Male-Male}     & AutoVC & 24.47               & 11.70 \\ \cline{2-4} 
                               & PPG-VC & 13.83               & 64.36 \\ \cline{2-4} 
                               & TTL-VC & 36.70               & 8.51  \\ \hline
Text-Male                      & MS-TTS & 25.00                  & 15.43 \\ \hline\hline 
\multirow{3}{*}{Male-Female}   & AutoVC & 3.85                & 28.85 \\ \cline{2-4} 
                               & PPG-VC & 0.96                & 67.31 \\ \cline{2-4} 
                               & TTL-VC & 54.81               & 0.96  \\ \hline
Text-Female                    & MS-TTS & 40.38               & 2.88  \\ \midrule[2pt]
\multirow{4}{*}{All}       & AutoVC & 15.59               & 27.24 \\ \cline{2-4} 
                               & PPG-VC & 11.55               & 59.78 \\ \cline{2-4} 
                               & TTL-VC & \textbf{43.05}      & \textbf{4.44}  \\ \cline{2-4} 
                               & MS-TTS & 29.81               & 8.54  \\ \bottomrule[2pt]
\end{tabular}
\label{table:bws}
\end{table}

\subsubsection{Subjective Evaluation}
\label{secex}

Subjective evaluations are performed through listening tests by human subjects.
Both AB and XAB preference tests were conducted to assess speech quality and speaker similarity, respectively.
In addition, to study listeners' preferences across all experimental systems, we further conducted best-worst scaling (BWS) and mean opinion score (MOS) tests~\cite{ccicsman2017sparse} on speaker similarity and speech quality, respectively. 20 samples were randomly selected from the converted samples of each experimental system and provided to 15 participants for all the tests~\footnote{The generated speech samples for all the source-target group pairs of each system are available at \url{https://arkhamimp.github.io/TTL-VC/}.}. All listeners are university students and staff members, where English is their official language of instruction.

\begin{figure*}
\centering
\includegraphics[width=160mm]{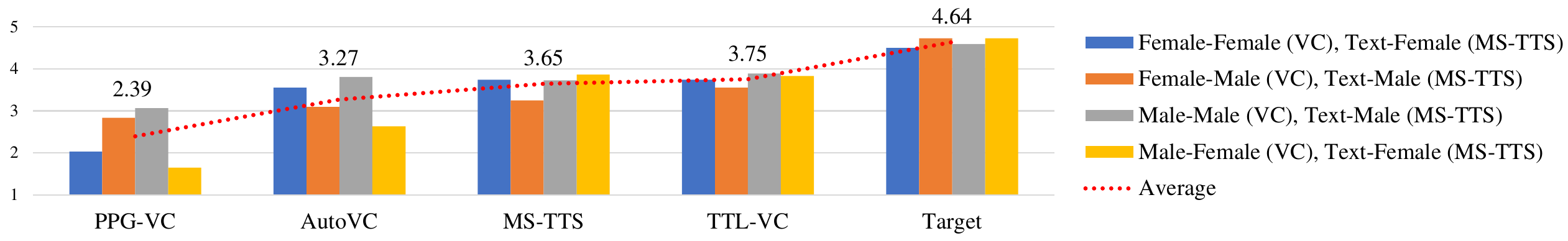}
\caption{MOS results of four models for four source-target groups and the average result.}
\label{fig:mos}
\end{figure*}

\begin{itemize}
    \item AB preference tests: A and B were speech samples randomly selected from different systems. The listeners were asked to choose the sample having higher quality and naturalness. We have three comparison sets, where 7 samples were evaluated by each listener. The results are illustrated in Fig. \ref{fig:ab}. All experiment results are reported with \textit{p}-value $<$ 0.01, while the ``TTL-VC vs. AutoVC" pair in female-female conversion is with \textit{p}-value = 0.017. Therefore, the performance gains by TTL-VC over other methods are statistically significant.  First, by comparing TTL-VC with PPG-VC, we observe that TTL-VC receives most of the preference votes for all source-target pairs. This suggests that context vectors outperform  PPG as linguistic features; Second, between TTL-VC and AutoVC, we observe that TTL-VC also consistently outperforms AutoVC across all source-target pairs. This confirms the advantage of context vectors over AE latent codes. Last, when we focus on the comparison between TTL-VC and MS-TTS, it is found that TTL-VC outperforms baseline MS-TTS in Female-Female, Female-Male and Male-Female conversions, while it performs slightly worse than MS-TTS in Male-Male conversions. Overall, TTL-VC outperforms other competing VC models and performs comparably with MS-TTS on average.

    \item XAB preference tests: X was the reference target speech sample; A and B were speech samples randomly selected from different systems. The listeners were asked to listen both samples, and choose the one more similar to the target speaker. For each comparison pair, 7 sets of samples were evaluated by each listener on average. The results are illustrated in Fig. \ref{fig:abx}. All experiments are reported with \textit{p}-values $<$ 0.01, while the ``TTL-VC vs. MS-TTS" pair in text-female conversion has \textit{p}-value = 0.033. This suggests that the performance gains are statistically significant. By comparing TTL-VC with the competing models, we observe that TTL-VC obviously outperforms PPG-VC and AutoVC for all speaker pairs in terms of speaker similarity; furthermore, TTL-VC and MS-TTS are on a par with each other.  
    
    \item BWS tests: We presented 4 samples of the same content from all four experimental systems to the listeners. The listeners were asked to pick the best and worst sample from the four. 20 scaling sets were evaluated by each listener. The detailed results are illustrated in Table \ref{table:bws}. Overall, TTL-VC receives the highest best votes (43.05\%) and the lowest worst votes (4.44\%) respectively. The results indicate that the performance of all four experimental systems is ranked in the descending order as: TTL-VC, MS-TTS, AutoVC and PPG-VC.
    
    \item MOS tests: The listeners were asked to rate the speech quality and naturalness of the converted speech on a 5-point scale. A higher score indicates better quality. For each experimental system, 20 samples were rated by each listener. The average results are illustrated in Fig. \ref{fig:mos}. We find that TTL-VC receives the highest MOS score (3.75) among all systems after the natural target speech (4.64), that slightly outperforms MS-TTS (3.65). With slight variations in different speaker pairs, we could rank the systems from high to low performance in the following order, TTL-VC, MS-TTS, AutoVC, and PPG-VC. The observation is consistent with that in other listening tests.

\end{itemize}

We observe that the proposed TTL-VC system clearly outperforms AutoVC and PPG-VC baselines in terms of speech quality, naturalness, and speaker similarity. In both MOS and BWS tests, it is encouraging to see that TTL-VC could succesully learn from TTS to achieve TTS quality without the need of text input at run-time inference.  


\section{Conclusion}
\label{secconc}
This paper presents a novel strategy for TTS-VC transfer learning, which has a simpler run-time inference system architecture, yet achieves consistently higher performance than other systems of similar architecture. We have demonstrated the effectiveness of the transfer learning algorithm and the system architecture. It is particularly encouraging that we observe that the proposed system not only provides high quality spectral mapping, but also prosodic rendering.


%

\section*{Acknowledgment}

This work was supported by the National Research Foundation, Singapore under its AI Singapore Programme (Award No: AISG-GC-2019-002) and (Award No: AISG-100E-2018-006), and its National Robotics Programme (Grant No. 192 25 00054), and by RIE2020 Advanced Manufacturing and Engineering Programmatic Grants A1687b0033, and A18A2b0046.
Yi Zhou is also supported by NUS research scholarship.

\ifCLASSOPTIONcaptionsoff
  \newpage
\fi

\newpage



\bibliographystyle{IEEEtran}
\bibliography{mybib}
%

%

\newpage
\begin{IEEEbiography}[{\includegraphics[width=1in,height=1.25in,clip,keepaspectratio]{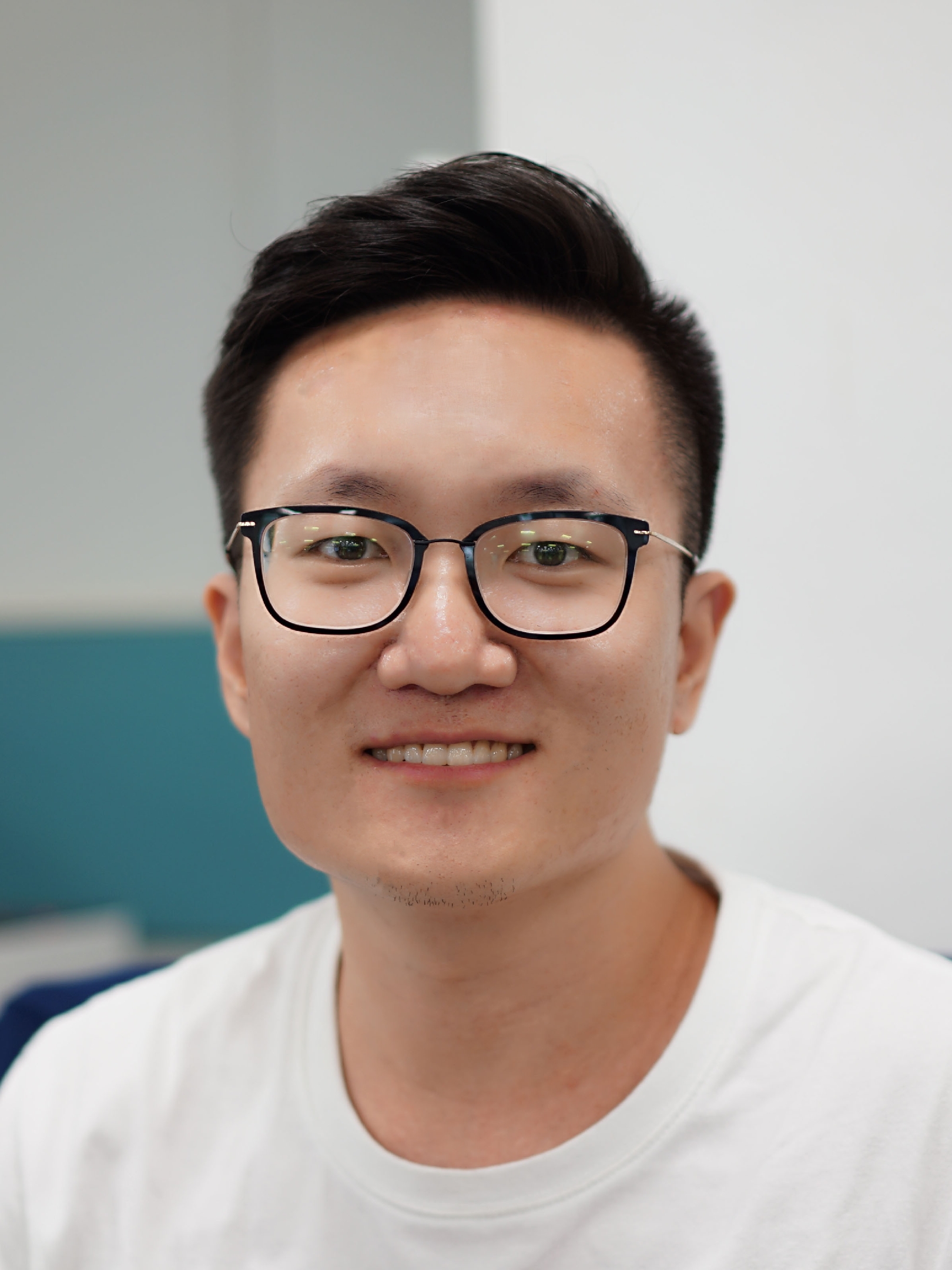}}]{Mingyang Zhang}
received the B.Eng. degree in Electrical Information Engineering from Nanjing University of Posts and Telecommunications, Nanjing, China, in 2014. He is currently working toward the Ph.D. degree in Key Laboratory of Underwater Acoustic Signal Processing of Ministry of Education, Southeast University, Nanjing, China. He is also with the Department of Electrical $\&$ Computer Engineering at the National University of Singapore for a research attachment. In 2018, he had a research internship in the National Institute of Informatics, Tokyo, Japan. His current research interests include speech synthesis and voice conversion.
\end{IEEEbiography}

\begin{IEEEbiography}[{\includegraphics[width=1in,height=1.25in,clip,keepaspectratio]{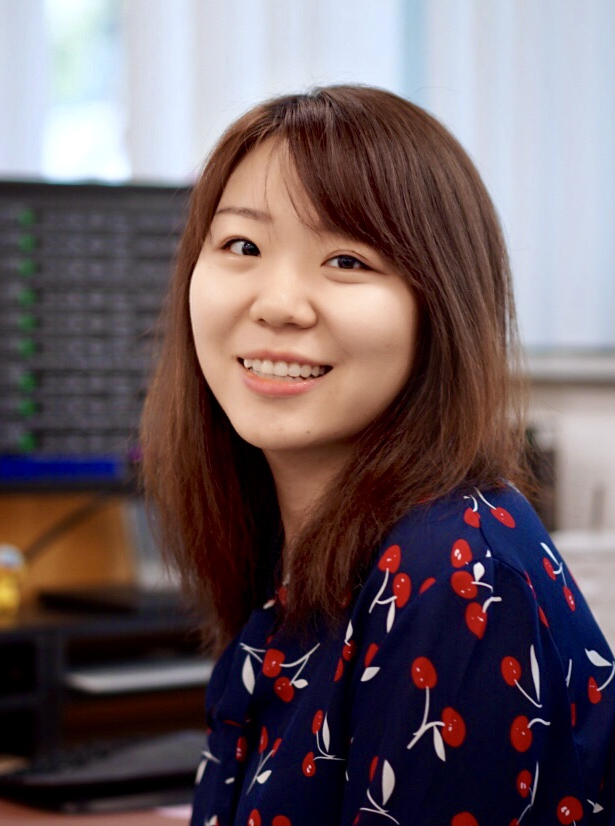}}]{Yi Zhou}
received the B.Eng. degree in Electrical and Electronic Engineering from Nanyang Technological University (NTU), Singapore, in 2015. She is currently a research scholar working toward the Ph.D. degree in Human Language Technology Lab, Department of Electrical $\&$ Computer Engineering at the National University of Singapore (NUS). Her research interest is cross-lingual voice conversion.

\end{IEEEbiography}


\begin{IEEEbiography}[{\includegraphics[width=1in,height=1.25in,clip,keepaspectratio]{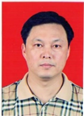}}]{Li Zhao}
received the B.E. degree from the Nanjing University of Aeronautics and Astronautics, China, in 1982, the M.S. degree from Suzhou University, China, in 1988, and the Ph.D. degree from the Kyoto Institute of Technology, Japan, in 1998. He is currently a Professor with Southeast University, China. His research interests include speech signal processing and pattern recognition.
\end{IEEEbiography}

\begin{IEEEbiography}[{\includegraphics[width=1in,height=1.25in,clip,keepaspectratio]{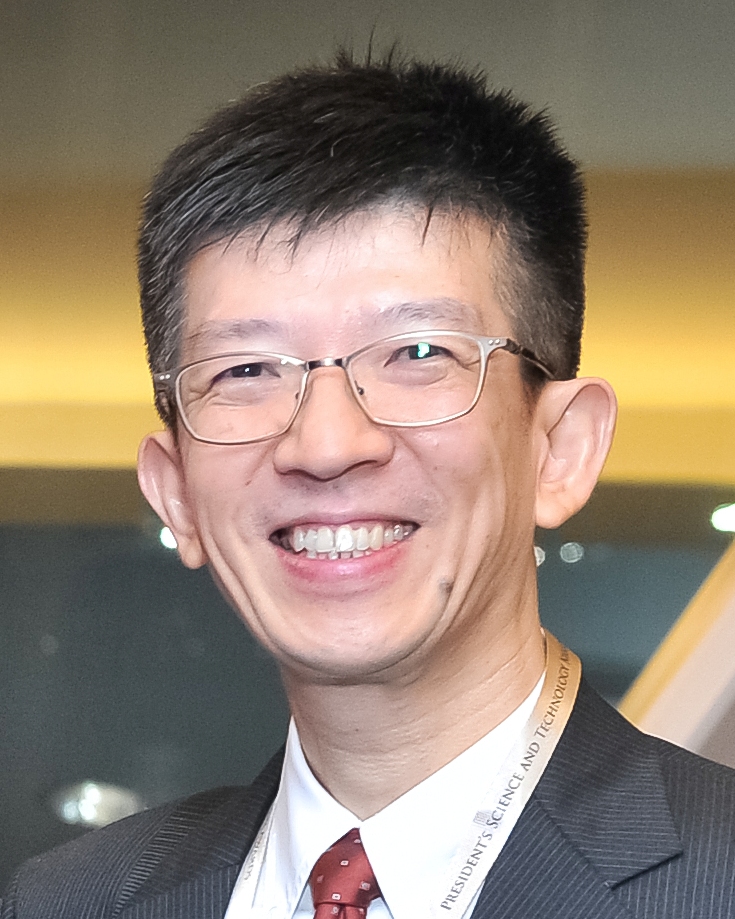}}]{Haizhou Li}
 (M’91-SM’01-F’14) received the B.Sc., M.Sc., and Ph.D degree in electrical and electronic engineering from South China University of Technology, Guangzhou, China in 1984, 1987, and 1990 respectively. Dr Li is currently a Professor at the Department of Electrical and Computer Engineering, National University of Singapore (NUS). His research interests include automatic speech recognition, speaker and language recognition, and natural language processing. Prior to joining NUS, he taught in the University of Hong Kong (1988-1990) and South China University of Technology (1990-1994). He was a Visiting Professor at CRIN in France (1994-1995), Research Manager at the Apple-ISS Research Centre (1996-1998), Research Director in Lernout \& Hauspie Asia Pacific (1999-2001), Vice President in InfoTalk Corp. Ltd. (2001-2003), and the Principal Scientist and Department Head of Human Language Technology in the Institute for Infocomm Research, Singapore (2003-2016). Dr Li served as the Editor-in-Chief of IEEE/ACM Transactions on Audio, Speech and Language Processing (2015-2018), a Member of the Editorial Board of Computer Speech and Language (2012-2018). He was an elected Member of IEEE Speech and Language Processing Technical Committee (2013-2015), the President of the International Speech Communication Association (2015-2017), the President of Asia Pacific Signal and Information Processing Association (2015-2016), and the President of Asian Federation of Natural Language Processing (2017-2018). He was the General Chair of ACL 2012, INTERSPEECH 2014 and ASRU 2019. Dr Li is a Fellow of the IEEE and the ISCA. He was a recipient of the National Infocomm Award 2002 and the President’s Technology Award 2013 in Singapore. He was named one of the two Nokia Visiting Professors in 2009 by the Nokia Foundation, and Bremen Excellence Chair Professor in 2019.
\end{IEEEbiography}




\end{document}